\documentclass[amsmath,amssymb,pra,aps,showpacs,superscriptaddress,twocolumn]{revtex4-1}
\pdfoutput=1
\usepackage{amsmath,amsfonts,amssymb,amsthm,graphics,graphicx,epsfig,bbm}
\usepackage[colorlinks=true,citecolor=blue,linkcolor=blue,urlcolor=blue]{hyperref}
\usepackage[usenames]{color}
\usepackage{graphicx}
\usepackage{subfigure}
\usepackage{amsmath}
\usepackage{epsfig}
\usepackage{dcolumn}
\usepackage{bm}
\usepackage{color}
\usepackage{epstopdf}
\usepackage{amssymb}
\usepackage{amstext}
\usepackage{latexsym}
\usepackage{hyperref}
\usepackage{amsfonts}
\usepackage{psfrag}
\usepackage{xcolor}
\usepackage[normalem]{ulem}
\usepackage{dsfont}
\usepackage{txfonts}

\graphicspath{{./Images/},{./ImagesAppendix/}}

\newcommand{\mean}[1]{\left\langle #1 \right\rangle}

\newcommand{\phiA}{\ensuremath{\phi_\text{A}}}
\newcommand{\phiB}{\ensuremath{\phi_\text{B}}}

\newcommand{\an}[2]{\ensuremath{\hat{#1}^{\protect\phantom{\dagger}}_{#2}}}
\newcommand{\cn}[2]{\ensuremath{\hat{#1}^\dagger_{#2}}}
\newcommand{\nn}[2]{\ensuremath{\hat{n}^{#1}_{#2}}}

\newcommand{\expU}[1]{\ensuremath{e^{#1}}}
\newcommand{\abs}[1]{\left|#1\right|}

\newcommand{\vc}[1]{\ensuremath{{\bf#1}}}
\newcommand{\diff}[1]{\ensuremath{\text{d}{\bf#1}}}
\newcommand{\op}[1]{\ensuremath{\hat{#1}}}

\newcommand{\subfigimg}[3][,]{%
	\setbox1=\hbox{\includegraphics[#1]{#3}}
	\leavevmode\rlap{\usebox1}
	\rlap{\hspace*{2pt}\raisebox{\dimexpr\ht1-0.5\baselineskip}{{\bfseries \large\textsf{#2}}}}
	\phantom{\usebox1}
}

\newcommand{\idg}[1]{{\bfseries #1)}}
\newcommand\numberthis{\addtocounter{equation}{1}\tag{\theequation}}

\newcommand{\rev}[1]{#1}
\newcommand{\revA}[1]{{#1}}
\begin{document}
	
\title{{Mesoscopic Vortex-Meissner currents in ring ladders}}

\author{Tobias Haug}
\affiliation{Centre for Quantum Technologies, National University of Singapore,
3 Science Drive 2, Singapore 117543, Singapore}
\author{Luigi Amico}
\affiliation{Centre for Quantum Technologies, National University of Singapore,
3 Science Drive 2, Singapore 117543, Singapore}
\affiliation{Dipartimento di Fisica e Astronomia, Via S. Sofia 64, 95127 Catania, Italy}
\affiliation{CNR-MATIS-IMM \&   INFN-Sezione di Catania, Via S. Sofia 64, 95127 Catania, Italy}
\affiliation{LANEF {\it 'Chaire d'excellence'}, Universit\`e Grenoble-Alpes \& CNRS, F-38000 Grenoble, France}
\affiliation{MajuLab, CNRS-UNS-NUS-NTU International Joint Research Unit, UMI 3654, Singapore}
\author{Rainer Dumke}
\affiliation{Centre for Quantum Technologies, National University of Singapore, 3 Science Drive 2, Singapore 117543, Singapore}
\affiliation{Division of Physics and Applied Physics, Nanyang Technological University, 21 Nanyang Link, Singapore 637371, Singapore}
\author{Leong-Chuan Kwek}
\affiliation{Centre for Quantum Technologies, National University of Singapore,
3 Science Drive 2, Singapore 117543, Singapore}
\affiliation{MajuLab, CNRS-UNS-NUS-NTU International Joint Research Unit, UMI 3654, Singapore}
\affiliation{Institute of Advanced Studies, Nanyang Technological University,
60 Nanyang View, Singapore 639673, Singapore}
\affiliation{National Institute of Education, Nanyang Technological University,
1 Nanyang Walk, Singapore 637616, Singapore}

\date{\today}

\begin{abstract}
Recent experimental progress have revealed Meissner and Vortex phases in low-dimensional ultracold atoms systems. Atomtronic setups can realize ring ladders, while explicitly taking the finite size of the system into account. This enables the engineering of quantized chiral currents and phase slips in-between them. We find that the mesoscopic scale modifies the current. Full control of the lattice configuration reveals a reentrant behavior of Vortex and Meissner phases. Our approach allows a feasible diagnostic of the currents' configuration through time of flight measurements.
\end{abstract}


 \maketitle

The response of quantum coherent systems to an external perturbation may be implying  subtle physical phenomena. A textbook example in this context  is provided by the Meissner effect.  Originally formulated in condensed matter, the Meissner effect explains how  a superconductor (a phase coherent system) thicker than the penetration depth expels magnetic fields. In a so called type II superconductor, however, the magnetic field can penetrate the superconductor, but it must be organized in a lattice of flux tubes (Abrikosov Vortex lattice)\cite{meissner1933neuer,rickayzen1969theory}. Indeed, such phenomenon is intimately related to the Anderson-Higgs mechanism in relativistic Yang-Mills theories, and  recent efforts  have been devoted to understand  whether Abrikosov vortices can occur in the Higgs field\cite{sudbo2013anderson}. 

Ultracold atoms confined in optical lattices allow us to study the above problem in a novel and fruitful way\cite{bloch2005ultracold}. Specifically,  ladder structures have been considered, in which ultracold bosonic atoms can tunnel between two one-dimensional chains. An artificial  gauge field  \cite{eckardt2005superfluid,dalibard2011colloquium,tung2006observation}
mimics the external magnetic field implied in the Meissner effect. In this context the perfect 'diamagnetism' arises from  currents flowing along the legs (chiral currents). A vortex phase may arise as a specific current configuration involving a flow of particles  along the  rungs of the ladder.
Recently, many experiments of such systems have been realized \cite{atala2014observation,mancini2015observation,stuhl2015visualizing,an2017direct,livi2016synthetic} and the theory has been studied in\cite{kardar1986josephson,granato1990phase,orignac2001meissner,crepin2011phase,tokuno2014ground,greschner2015spontaneous,piraud2015vortex,kelecs2015mott,di2015meissner,orignac2016incommensurate,guo2016geometry,guo2017dissipatively}. It was understood, in particular, that the commensurate Meissner state undergoes a quantum phase transition to an incommensurate Vortex lattice for open boundary ladders. 
Recently, a very interesting Vortex-charge duality was shown \cite{greschner2017vortex}.

To study the problem, here we are inspired by Atomtronics: optical circuits of very different spatial shapes and intensity for manipulation of cold atoms\cite{seaman2007atomtronics,Amico_Atomtronics}.
 Atomtronic quantum technology aims at devising a circuitry of a new type with atomic currents. 
At the same time, with closed confinements,  it can enable a new platform for cold-atoms current-based quantum simulators to explore quantum many-body phases\cite{amico2005quantum}.
The Meissner/Vortex transition described above provides a striking example in which Atomtronics can demonstrate its full potential.  In this paper,  we study the physics implied by Meissner and Vortex phases at the mesoscopic scale in ladders with closed boundary conditions. We shall see that quantum phase slips\cite{matveev2002persistent,rastelli2013quantum,pop2010measurement,astafiev2012coherent,roscilde2016} and flux quantization\cite{wright2013driving} play important roles in the physics of these system. The physical platform of the system will be provided by a specific  Atomtronic set-up made of  two  coupled bosonic condensates confined in 'on-top' ring-shaped optical potentials\cite{li2008real, amico2014superfluid, aghamalyan2013effective}. With our approach, we will explain the following: First, how  the  configuration of the current in the two phases are related to persistent currents flowing in the ring condensates. And secondly, how to measure the phases with absorption imaging of the condensate. 
\revA{Despite its noteworthiness,  the qubit dynamics  encoded into the system is discussed in  the   Appendix \ref{qubitdynamics} so as not to distract the reader from the main theme of the manuscript.}

{\em Model--}
\begin{figure}[htbp]
	\centering
	\subfigure{\includegraphics[width=0.48\textwidth]{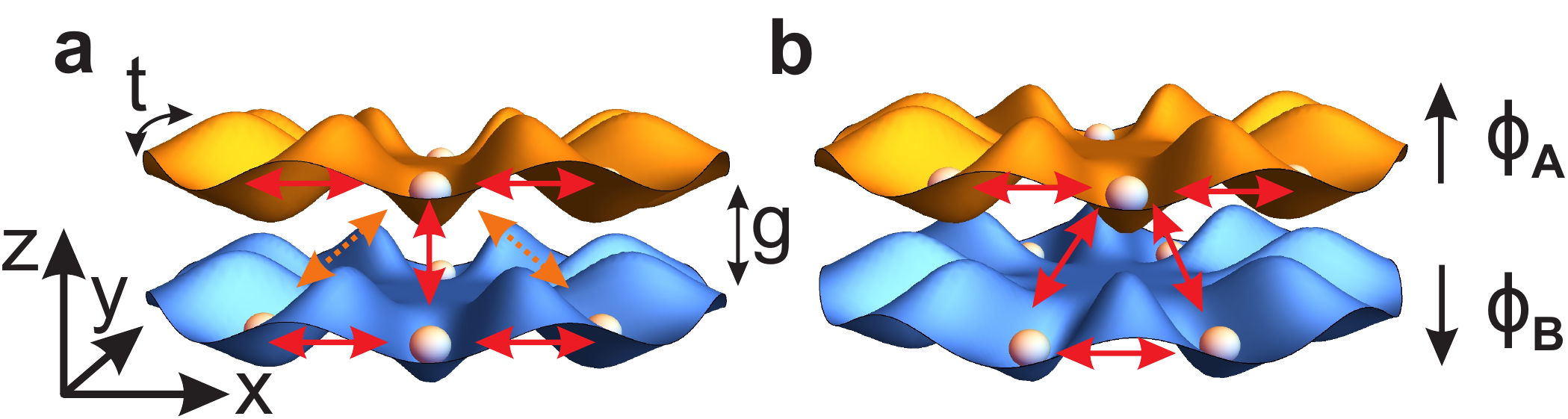}}
	\subfigure{\includegraphics[width=0.45\textwidth]{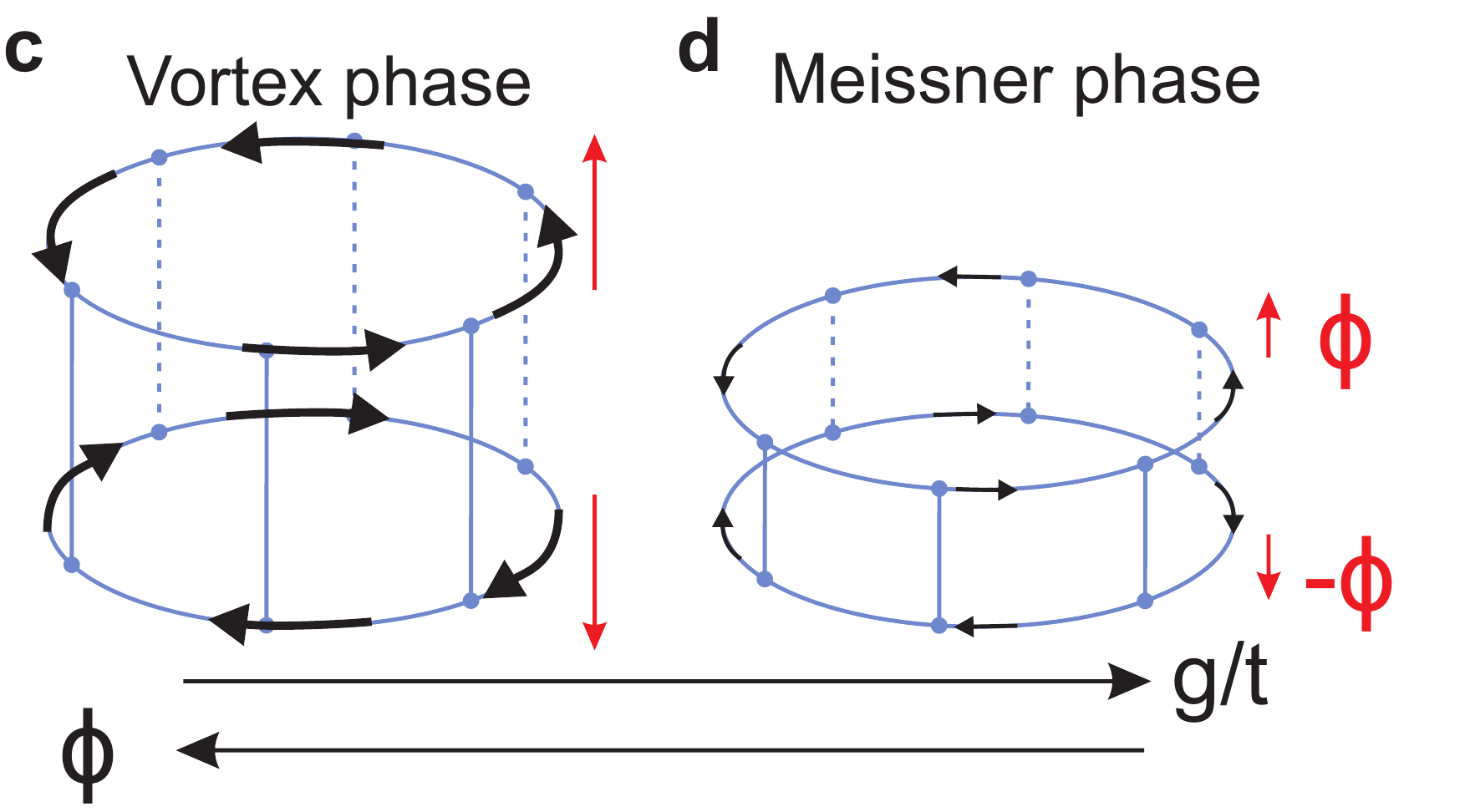}}
	\caption{Two on-top ring lattice potentials loaded with bosons. Atoms can tunnel to the nearest neighbors in the ring as well as to sites on the other ring. Rings can be twisted in respect to each other to shift relative position of sites. Upper and lower ring are threaded by a flux. In {\bfseries a)}, ring sites are aligned and inter-tunneling (denoted by red arrows) occurs only between adjacent sites. \rev{Optionally, diagonal inter-ring tunneling (dashed orange arrow) can be introduced.} In {\bfseries b)}, rings are twisted relative to each other, such that a site on ring A couples equally to two sites on ring B. The result is a triangular lattice. The red arrows indicate the possible tunneling between lattice sites. {\bfseries c)} Vortex-Meissner phase characterized by the quasi-momentum in a configuration with opposite flux in each ring. \revA{The arrows denote the direction and average value of the quasi-momentum.} For weak inter-ring coupling $g/t$ or a high flux $\phi$: Opposite, large average quasi-momentum in each ring (Vortex phase). {\bfseries d)} When inter-ring coupling is increased or flux is decreased: \revA{Zero quasi-momentum for non-interacting rings or a small residual quasi-momentum with interaction} (Meissner phase).}
	\label{Sketch}
\end{figure}
Two rings A and B with each $L$ sites are coupled via the rungs. A sketch of the model is shown in Fig.\ref{Sketch}a. The Hamiltonian $\mathcal{H}=\mathcal{H_\text{A}}+\mathcal{H_\text{B}}+\mathcal{H_\text{I}}$ is given by  
\begin{align*}
\mathcal{H_\text{A}}={}&\sum_{m=1}^{L}\left(-t\expU{i\phiA}\cn{a}{m}\an{a}{m+1}+\text{h.c.}\right)+\frac{U}{2}\nn{a}{m}(\nn{a}{m}-1)\\
\mathcal{H_\text{B}}={}&\sum_{m=1}^{L}\left(-t\expU{i\phiB}\cn{b}{m}\an{b}{m+1}+\text{h.c.}\right)+\frac{U}{2}\nn{b}{m}(\nn{b}{m}-1)\label{Hamilton}\numberthis\\
\mathcal{H_\text{I}}={}&\sum_{m=1}^{L}-g\left[(1-w)\cn{a}{m}\an{b}{m}\phantom{\frac{w}{2}}\right.\\
&\left.+\frac{w}{2}(1+\gamma)\cn{a}{m}\an{b}{m+1}+\frac{w}{2}(1-\gamma)\cn{a}{m}\an{b}{m-1}\right]+\text{h.c.} \; .
\end{align*}
Operators $\an{a}{m}$ ($\cn{a}{m}$) and $\an{b}{m}$ ($\cn{b}{m}$) destroy (create) a boson at site $m$ in ring A and B respectively and ${\nn{a}{m}\doteq \cn{a}{m} \an{a}{m}}$ and ${\nn{b}{m}\doteq \cn{b}{m} \an{b}{m}}$ are the particle number operators. The parameter  $t$ corresponds to the intra-ring coupling, $g$ to the inter-ring coupling; \rev{ring-twist $w$ and $\gamma$ are parameters accounting for different coupling geometries between the two rings} (as described in detail later on), $U$ the on-site interaction and $L$ the number of sites per ring. The operators are constrained by periodic boundary conditions for each ring ${\alpha=\{a,b\}}$ with ${\an{\alpha}{L+1}=\an{\alpha}{1}}$. An artificial gauge field $\vc{A_\alpha}$ is introduced through the Peierls substitution 
${t\rightarrow t\expU{i\phi_\alpha}}$
, where ${\Omega_\alpha=\phi_\alpha L=\frac{q}{\hbar}\oint_C\vc{A_\alpha}\diff{\vc{r}}}$ is the total flux threading each ring, and $\phi_\alpha$ the phase acquired when tunneling between neighboring sites. 
The artificial gauge field can be created in different ways\cite{dalibard2011colloquium,goldman2014light}. 

The phase term can be moved to the inter-ring coupling with the transformation $\an{\alpha}{m}\rightarrow \an{\alpha}{m}\expU{-im\phi_\alpha}$\cite{osterloh2000exact,amico1998one}.

The phase winding in the ring is quantized to an integer winding number ${n=\frac{1}{2\pi}\oint_C\nabla\Theta\diff{r}}$ \cite{kashurnikov1996supercurrent} and represents a topological quantity which is associated with the persistent current\cite{ragole2016interacting}. 

\rev{We now return to the ring-twist parameter $w$ and $\gamma$ mentioned in Eq.\ref{Hamilton}.
These parameters are introduced in view of the new perspectives opened up by the  Atomtronics quantum technology. We propose two ways to implement this parameter in experiment: Either by engineering diagonal inter-ring tunneling (${\gamma=0}$, see Fig.\ref{Sketch}a with dashed coupling) or angular shift of the lattice sites of ring A relative to ring B (${\gamma=1}$). The minimal twist ${w=0}$ is a simply connected ladder (see Fig.\ref{Sketch}a without dashed coupling). For the first method the maximal twist ${w=0.5}$ is achieved by configuring equal inter-ring tunneling rates for diagonal and direct coupling, while for the second configuration it realizes a triangular lattice (see Fig.\ref{Sketch}b and \cite{mishra2013quantum,anisimovas2016semisynthetic,longhi2014aharonov}). Details on the implementation of the diagonal tunneling (${\gamma=0}$) is found in Appendix \ref{AppendixDiag} and for the relative angular shift (${\gamma=1}$) in Appendix \ref{AppendixRotation}.}

{\it Configuration of currents--}
We consider opposite flux in each ring ${\phi=\phiA=-\phiB}$, which features the Meissner-Vortex transition mentioned in the introduction. 
We characterize this phase by looking at the quasi-momentum $k$. The number of particles with quasi-momentum $k$ in ring $\alpha$ is given by
\begin{equation}
\nn{}{\alpha}(k)=\cn{\alpha}{k}\an{\alpha}{k}=\frac{1}{L}\sum_{n,n'}\expU{ik(n-n')}\cn{\alpha}{n}\an{\alpha}{n'}\;.
\end{equation}
We define the difference of center-of-mass quasi-momentum in ring A and B (called chiral momentum from now on)
\begin{equation}\label{Eqchiralmomentum}
{K_\text{c}=\sum_kk\mean{\nn{}{A}(k)-\nn{}{B}(k)}/N_\text{p}}\;,
\end{equation}
where $N_\text{p}$ is the number of particles and ${k\in (-\pi,\pi]}$. 
The quasi-momentum $k$ is quantized as ${k_n=2\pi n/L}$, where $n$ is the phase winding number. 
The phase winding in the ring can be measured via time-of-flight expansion\cite{wright2013driving,eckel2014interferometric,aghamalyan2015coherent}. 
Chiral momentum $K_\text{c}$ is the order parameter of the Vortex-Meissner phase, as for a non-interacting system it is zero in the Meissner phase (as both rings have the same quasi-momentum distribution), and non-zero in the Vortex phase with ${K_\text{c}\sim\phi}$ when the inter-ring coupling $g/t$ is small. We find that with interaction and a finite number of lattice sites, the chiral momentum of the Meissner phase becomes non-zero, but is small compared to the Vortex phase.

A sketch of the configuration and currents is shown in Fig.\ref{Sketch}c,d. For small inter-ring coupling $g/t$ or large flux per site $\phi$, the two ring condensates are effectively decoupled and rotate independently of each other with the external flux, resulting in a large chiral momentum. The Vortex phase is defined by the ground state being the  superposition of two degenerate states with counter-propagating quasi-momentum\cite{wei2014theory,tokuno2014ground}. \revA{As we impose periodic boundary conditions, a Vortex lattice structure is not visible in the density along the legs as it has been seen in previous studies with open boundary conditions. Instead, a 'vortex fluid' emerges as specific spatial features in higher-order density-density and current-current correlations.} 

In the other limit with large $g/t$ and small $\phi$ (Meissner phase), the inter-ring tunneling locks the phase of the bosons in both rings. As a result, the bosons coherently cooperate to make the chiral momentum vanish. 

Note that in Refs \cite{orignac2001meissner,cha2011two,atala2014observation} the Vortex-Meissner phase is characterized by the chiral current, which is the difference of the average currents in ring A and B. 
Then, the Meissner phase has maximal chiral current and Vortex diminishing chiral current. 
This is the natural definition in previous studies, where the system under study is a ladder with open boundaries. The flux affects the particle hopping along the rungs of the ladder and there is a protocol to measure the current in-situ\cite{atala2014observation}.
The momentum distribution of time-of-flight reflects this as well, with two peaks (proportional to $\pm\phi$) in Meissner phase, and four peaks (close to zero momentum) in Vortex phase. 

However, for the ring ladder, the flux can be induced along the rings (legs of the ladder), which corresponds to a different gauge. \revA{Different gauge choices may lead to  different time-of-flight images\cite{moller2010condensed,lin2011synthetic,kennedy2015observation}.  
For open ladders, the effect of the gauge transformation in the time-of-flight  expansion was studied in \cite{greschner2016symmetry}. Below, we  shall see how the ring ladder (Eq. \ref{Hamilton}) can be read-out using time-of flight images.}

We choose the chiral momentum  Eq. (\ref{Eqchiralmomentum}) as the main observable. In a ring setup, the canonical momentum is quantized in terms of the phase winding number $n$: $k=2\pi n/L$. The phase winding in rings has been studied extensively\cite{wright2013driving,eckel2014interferometric,aghamalyan2015coherent}. 
We can translate between the chiral current and the chiral momentum. \revA{In particular, the relation between our quasi-momentum distribution (in our gauge) and the chiral current corresponding to the rung gauge (as has been used in most of the previous studies) reads as}
\begin{equation}\label{Eqchiralcurrent}
\mean{j_\text{c}}=\frac{2t}{L}\sum_k\mean{\nn{}{A}(k)\sin(\phi-k)-\nn{}{B}(k)\sin(-\phi-k)}\;.
\end{equation}

\begin{figure}[htbp]
	\centering
	\subfigimg[width=0.24\textwidth]{a}{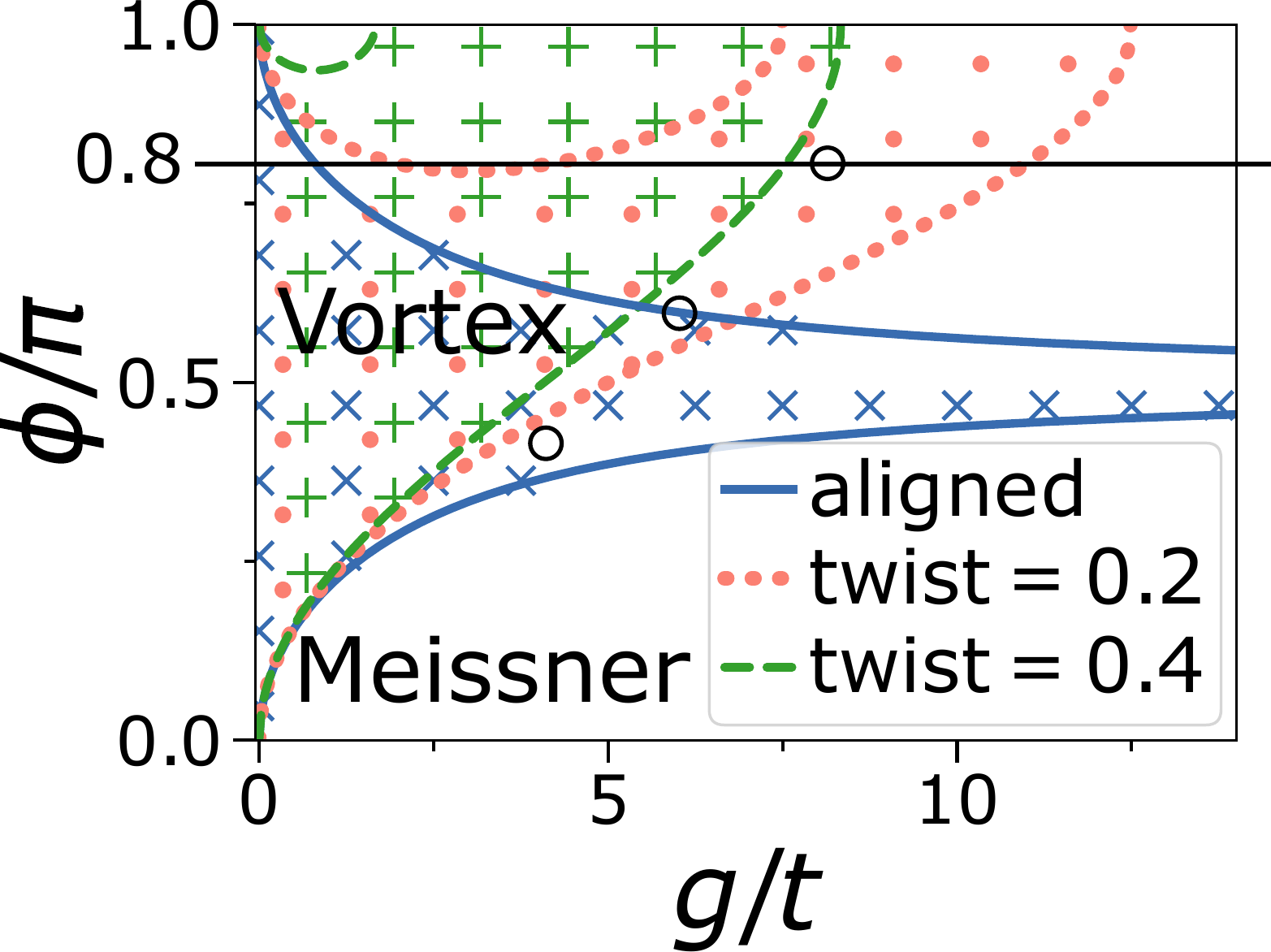}\hfill
	\subfigimg[width=0.24\textwidth]{b}{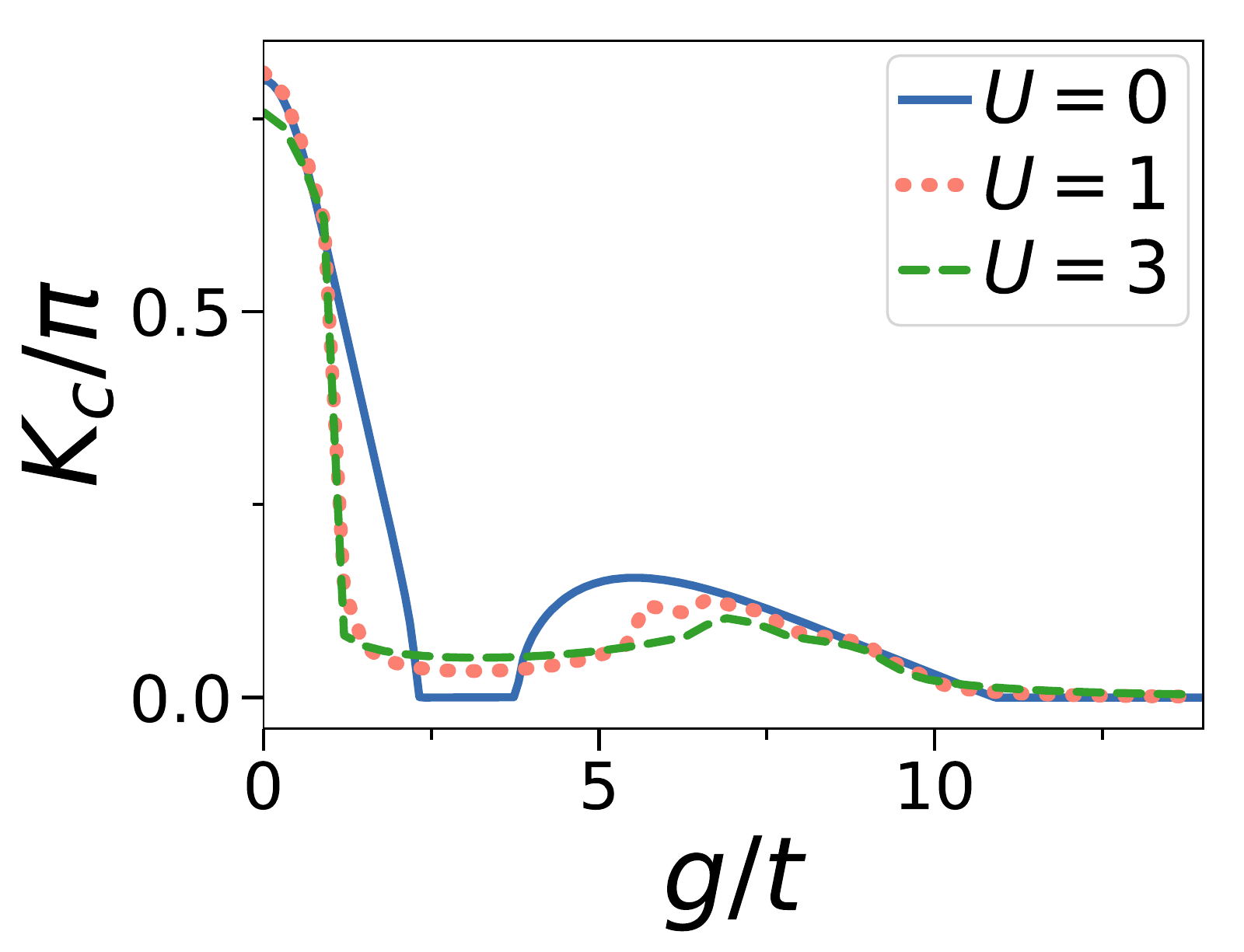}
	
	\caption{ {\bfseries a)}  \rev{Noninteracting} Meissner-Vortex phase diagram showing flux per site $\phi$ against inter-ring coupling $g/t$ for three values of ring-twist $w$ and \rev{for ${\gamma=1}$}. The Vortex phase area is denoted by the dots/crosses. \revA{There is a reflection symmetry at ${\phi=\pi}$.} Diagram depends strongly on ring-twist for ${\phi>0.5\pi}$. It shows a reentrant behaviour between the phases along the $g/t$ axis for intermediate ring-twist and ${\phi\approx0.8\pi}$. Line and open circle reference the cut through the phase diagram shown in Fig.\ref{CurrentChiralTwist}.
	\rev{{\bfseries b)} Chiral momentum $K_\text{c}$ against $g/t$ for different values of interaction $U/t$ for twist ${w=0.2}$. The corresponding cut through the phase diagram for flux $\phi=0.8\pi$ is indicated as black line in a).  \revA{For ${U=0}$ $K_\text{c}$ is zero in the Meissner phase, whereas for ${U>0}$ $K_\text{c}$ is non-zero as other quasi-momentum modes are excited by scattering.} The small steps in the profile for ${U>0}$ result from the quantized phase winding for small rings. The Vortex phase is characterized by an increased chiral momentum. Numerical result with 12 sites per ring and 6 particles in total.}	}
	\label{CritFlux}
\end{figure}

{\em Vortex/Meissner phases--}
First, we consider the Meissner-Vortex transition for zero interaction $U$. This allows us to consider a very large number of sites $L$ so that the effect of the periodic boundary conditions can be disregarded. \rev{We investigate ${\gamma=1}$, similar results for ${\gamma=0}$ are presented in Appendix \ref{AppendixDiag}.} The phase diagram \rev{for no interaction} is presented in Fig.\ref{CritFlux}a. 
We find two areas with Meissner phases, separated by the Vortex phase (dotted/crossed area). The area of the upper Meissner phase becomes smaller with increasing twist $w$. \revA{The lower part of the Meissner phase has a positive chiral current and ordering characterized by  $k=0$, while the upper part displays negative chiral current and  ordering $k=\pi$. 
For the rectangular ladder (${w=0}$), we find a reflection symmetry at ${\phi=\pi/2}$, which is broken for $w\neq0$. By inspection, we find that for ${g/t\ll1}$ or ${\phi\ll\pi/2}$ the phase diagram is nearly independent of ring-twist $w$.  Remarkably, by modifying only $w$ it is possible to switch the phase of the system.
In particular, for ${\pi>\phi>\pi/2}$ and ${w\approx0.2}$ a patch of Meissner phase is enclosed by the Vortex state}. This area becomes smaller with increasing $w$. {\it Therefore a reentrant behaviour is found  by increasing $g$ for ${\phi\approx 0.8\pi}$.} \rev{It persists for non-zero interaction as seen in Fig.\ref{CritFlux}b. For ${\gamma=1}$ and ${\phi=\pi}$ the dispersion is flat at the reentrance.}

\begin{figure}[htbp]
	\centering
	\subfigimg[width=0.24\textwidth]{a}{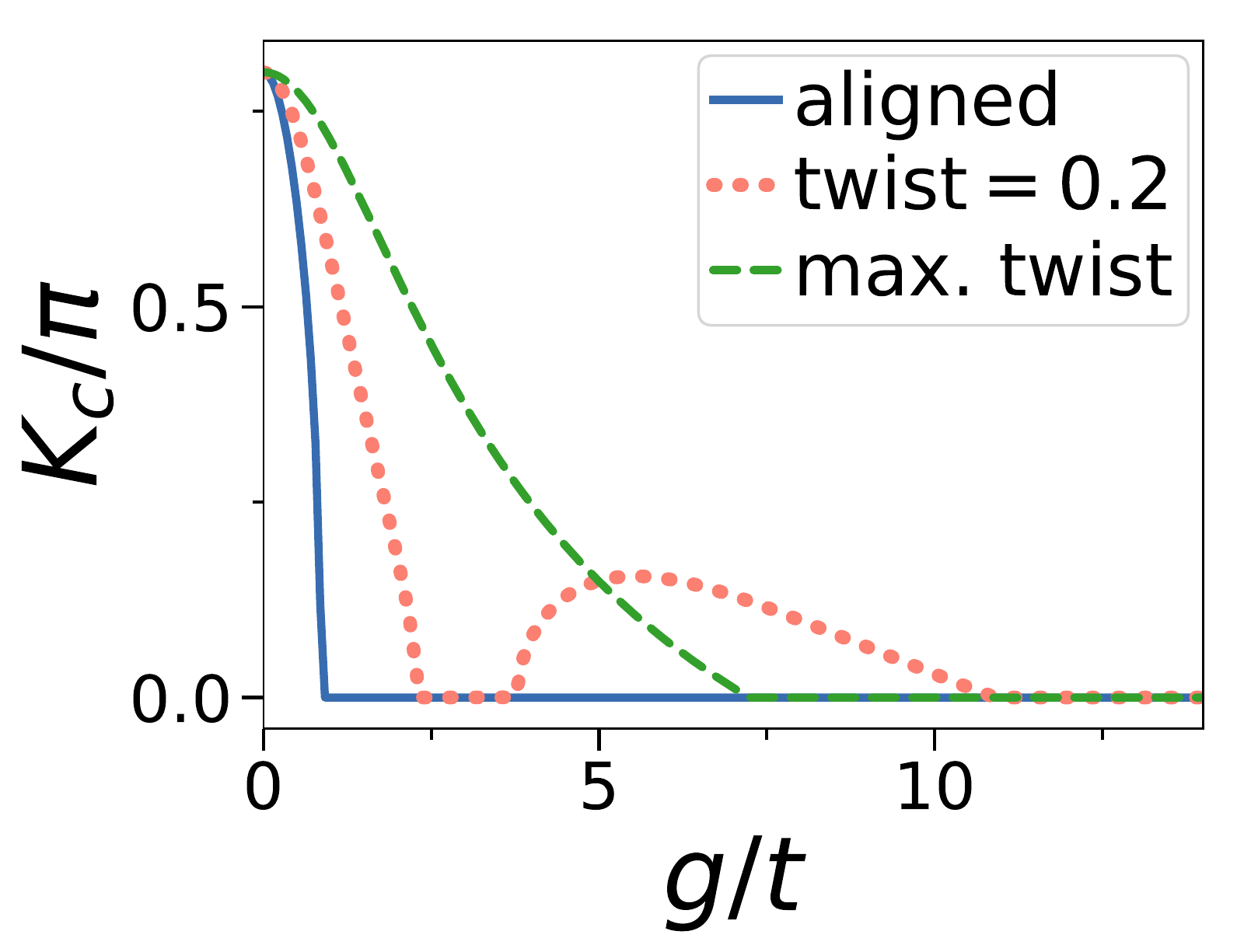}\hfill
	\subfigimg[width=0.24\textwidth]{b}{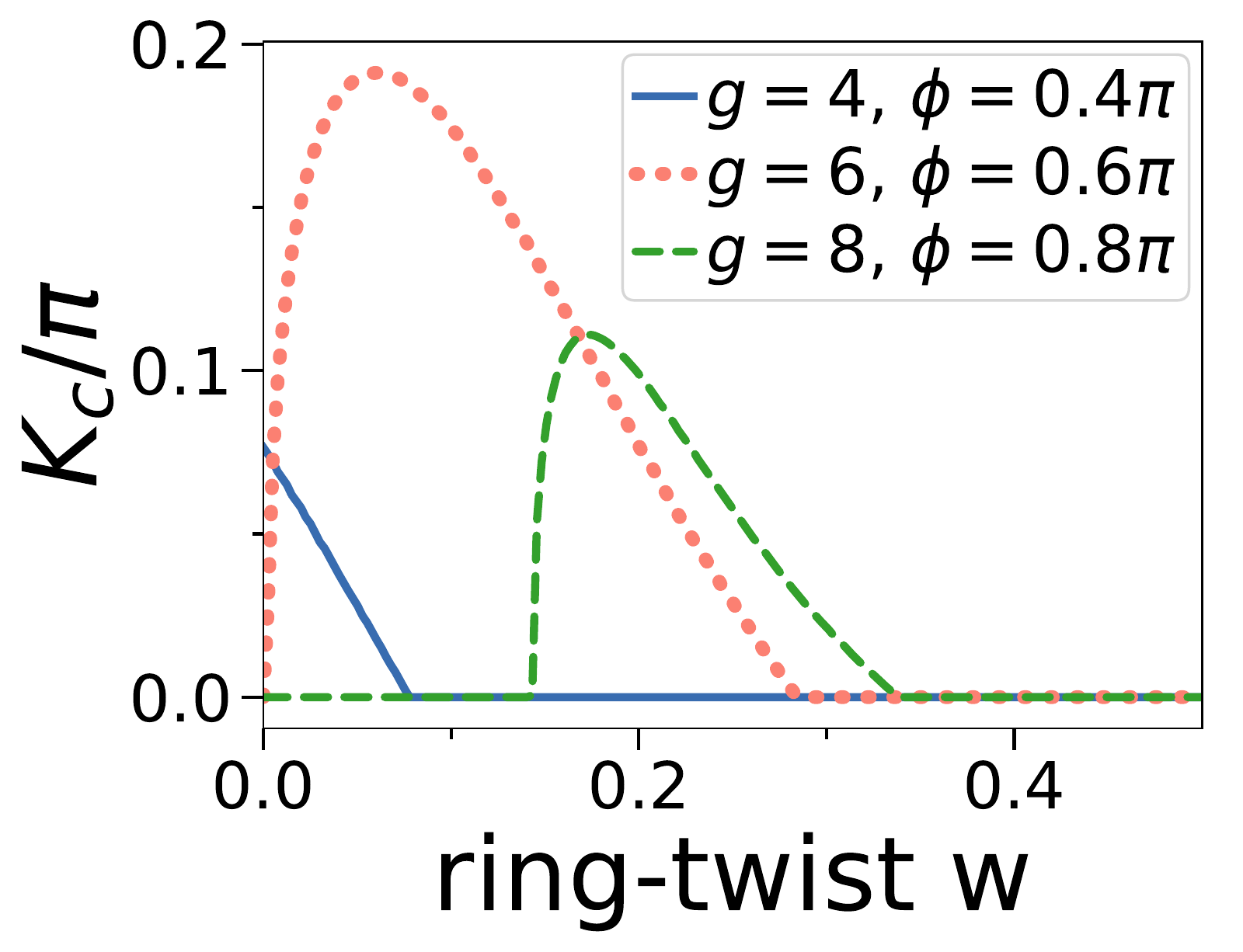}
	
	\caption{{\bf a)} Chiral momentum $K_\text{c}$ for a large number of sites and zero interaction plotted against inter-ring coupling $g/t$ and flux per site ${\phi=0.8\pi}$ \rev{for ${\gamma=1}$}. The corresponding cut through the phase diagram is shown as line in Fig.\ref{CritFlux}. For ${w=0.2}$, reentrance of the chiral momentum is observed. 
	{\bf b)} $K_\text{c}$ plotted against ring-twist for different values of $g/t$ and $\phi$. The open circles in Fig.\ref{CritFlux} show the corresponding positions in the phase diagram.}
	\label{CurrentChiralTwist}
\end{figure}

We plot the chiral momentum $K_\text{c}$ for various cuts through the phase diagram in Fig.\ref{CurrentChiralTwist}. In Fig.\ref{CurrentChiralTwist}a, we plot $K_\text{c}$ against inter-ring coupling $g$ for different ring-twist configurations and ${\phi=0.8\pi}$. While in a rectangular ladder it decreases very quickly, for maximal ring-twist the Vortex phase with non-zero $K_\text{c}$ extends to nearly ${g/t=8}$. This is due to destructive interference of inter-ring tunneling in triangular configuration. For intermediate ring-twist ${w=0.2}$ the reentrance is observed: With increasing inter-ring coupling, the chiral momentum vanishes at ${g/t\approx 3}$. $K_\text{c}$ resurfaces at intermediate couplings and then it is suppressed  for larger $g/t$.

In Fig.\ref{CurrentChiralTwist}b $K_\text{c}$ is plotted against different degrees of ring-twist from rectangular (${w=0}$) to triangular (${w=0.5}$) configurations. Depending on inter-ring coupling and flux, the ring-twist $w$ can control the chiral momentum as well as  it can drive the Meissner  to Vortex phase. For ${g/t=8}$ and ${\phi=0.8\pi}$ increasing $w$ reveals again the characteristic reentrant behavior of the Meissner-Vortex-Meissner transition. 

\begin{figure}[htbp]
	\centering
	\subfigimg[width=0.235\textwidth]{a}{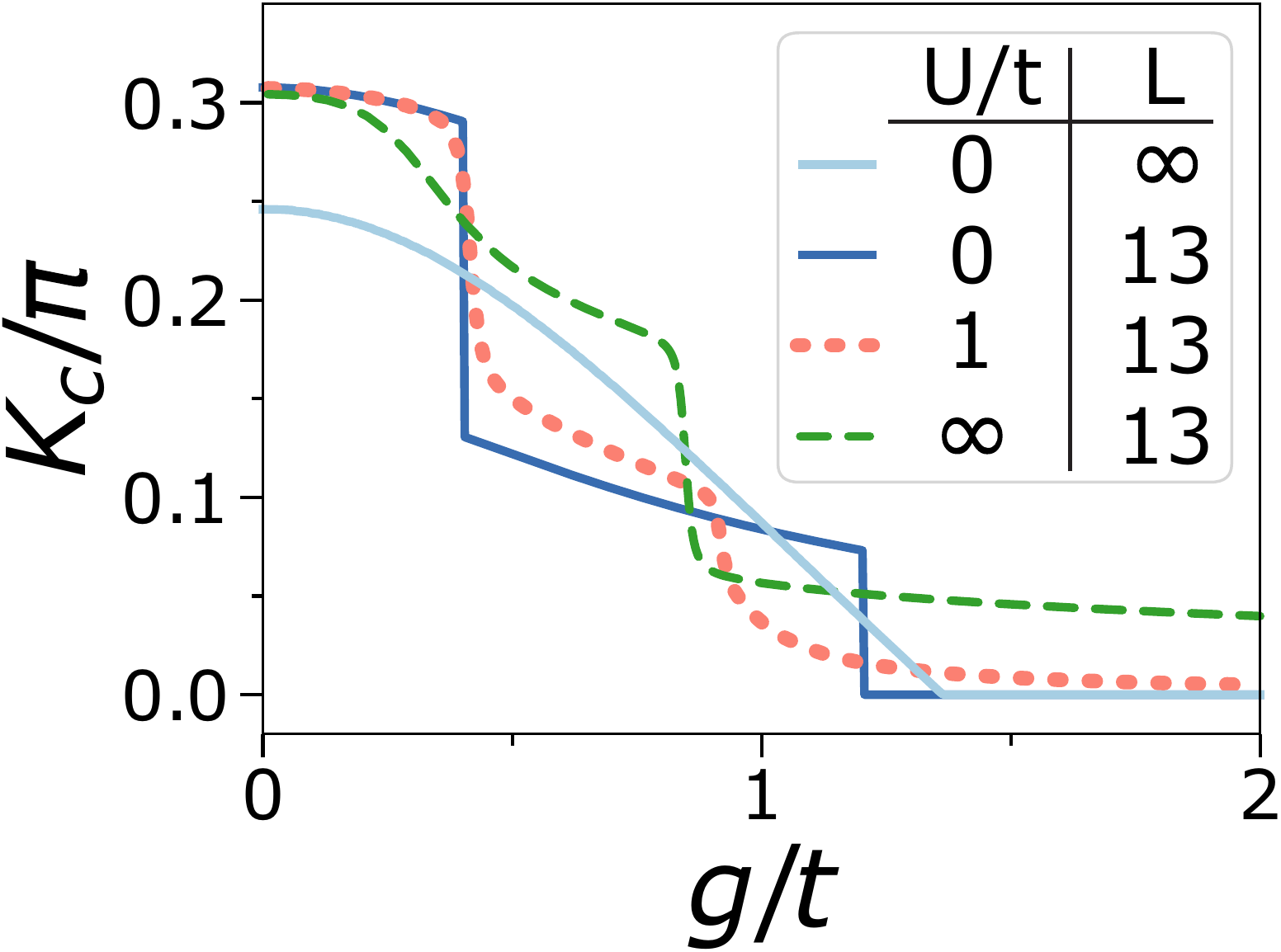}\hfill
	\subfigimg[width=0.245\textwidth]{b}{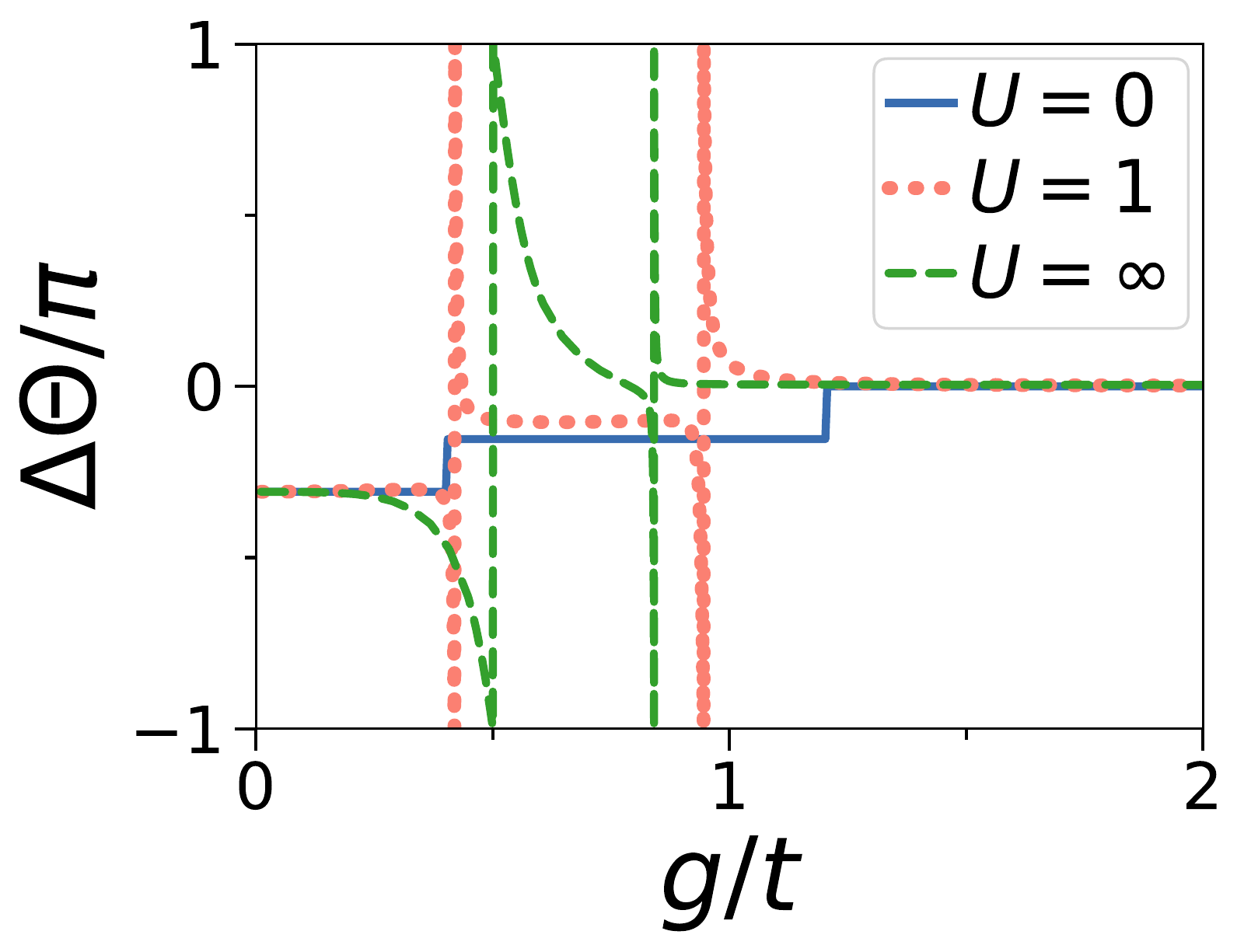}
	\caption{{\bf a)} Chiral momentum $K_\text{c}$ plotted against inter-ring coupling $g/t$ for different values of interaction $U/t$. Numerical result with ${L=13}$ sites per ring, 6 particles in total, ${\gamma=1}$, ${w=0}$, and a total flux ${\phiA L=-\phiB L=3.2\pi}$ (${\phiA=0.246\pi}$). The solid, light blue curve shows the chiral momentum for a large number sites. For small $g/t$, the $K_\text{c}$ depends quadratically on inter-ring coupling. For ${L=13}$ and ${g=0}$, the groundstate has 2 (-2) phase windings in ring A (B) and is in the Vortex phase. Chiral momentum and phase winding decrease with $g/t$ in two steps to nearly zero, which corresponds to the Meissner phase. With increasing $U/t$, second step appears at smaller $g/t$ and transition smears out.
	{\bf b)} Phase jump of two point correlations $\Delta\Theta={\text{arg}(\cn{a}{1}\an{a}{7})-\text{arg}(\cn{a}{1}\an{a}{6})}$ in ring A. Whenever the phase jump becomes $\pi$ a transition between two different phase winding states occurs.
}
	\label{CurrentChiralU}
\end{figure}
Next, we want to investigate the effect of a finite number of lattice sites $L$ on the ground state chiral momentum for zero ring-twist $w$. 
For small $L$, $K_\text{c}$ is not a continuous function anymore, but changes as a stair case for ${U=0}$--Fig.\ref{CurrentChiralU}a.  
In particular, we note that an offset appears in $K_c$ for finite $L$ (similarly in the chiral current $\langle j_c \rangle$). {\it Such offset has genuine mesoscopic origins tracing back to  persistent current flowing,  and quasi-momentum quantization in the rings}.
 When increasing $g/t$, we observe sharp transitions. Such behaviour arises because of  jumps between different values of quasi-momentum corresponding, in turn, to different winding numbers (a similar effect was evidenced in the Josephson current through  a Luttinger liquid ring \cite{fazio1995josephson}). In between steps, $K_\text{c}$ does not define a plateau (strictly constant value between two steps), but decreases monotonously. The reason is the following: For small inter-ring coupling, each ring carries phase windings with opposite values (e.g. ${n=2}$ in ring A, ${n=-2}$ ring B), well localized in each ring. \revA{By increasing the inter-ring coupling, the phase windings between the rings are mixed up, resulting in a suppression of the chiral momentum.}
In the Meissner phase, the phase winding is completely delocalized, thereby suppressing $K_\text{c}$.
A similar behaviour is seen in the chiral current as outlined in Appendix \ref{chiralcurrentApp}.
The scenario emerging from the above results indicates that the {\it Meissner-Vortex phases transition in our system occurs because of quantum  phase slips}\cite{matveev2002persistent}. In the Vortex phase, where the rings are effectively decoupled, upper and lower condensate have opposite phase winding. When increasing the inter-ring coupling, quantum phase slips occur between the opposing rotation states, which eventually cancel the phase winding difference in each ring to an average of zero.

Now we discuss the effect of non-zero interaction on  $K_\text{c}$.  For all values of interaction, a monotone decrease of the chiral momentum with $g/t$ is observed, with two sharper drops now instead of the discontinuous jump. As explained above, the sharp drops mark the transition between rotation states with different integer phase windings. The smooth transition heralds the appearance of superposition states of different topological phase winding numbers. 
\revA{To pinpoint at what value of inter-ring coupling quantum phase slips occur that change the winding number of the groundstate, we refer to the phase of the two point correlation function  $\Delta\Theta={\text{arg}(\cn{a}{1}\an{a}{\lfloor L/2+1\rfloor})-\text{arg}(\cn{a}{1}\an{a}{\lfloor L/2-1\rfloor})}$ in ring A.
Indeed, this phase changes by $\Delta\Theta=2\pi(1-1/L)$ whenever the ground state switches to a different phase winding.  Fig.\ref{CurrentChiralU}b shows the phase jump of the two point correlations. This effect is best visible for an odd number of ring sites.} The actual value of $\Delta\Theta$ is determined by the degree of superposition of two phase windings\cite{danshita2010accurate}.  

The specific value of ${g/t}$, at which the steps appear, shifts with increasing interaction. However, we observe that the second step changes more with on-site interaction. The second step shifts from about ${g/t\approx1.2}$ (${U=0}$) to ${g/t\approx0.84}$ (${U=\infty}$). Thus, the Meissner phase (minimal value of chiral momentum) appears at a lower values of coupling $g/t$ with increasing interaction (a similar effect was noticed  in \cite{piraud2015vortex} also for ladders with open boundaries). 
We explain this phenomenon as following: The Meissner state has a small chiral momentum, with largest contribution with zero phase winding. However, the Vortex state consists of opposite, non-zero phase winding flows in ring A and B. These adverse flows have additional scattering mechanisms compared to the zero winding case, making the Vortex phase energetically more unfavorable compared to the Meissner phase with increasing interaction.
We observe that \revA{in the Meissner phase ($g/t$ large)} the absolute value of chiral momentum increases with interaction. We attribute this to scattering into higher momentum modes.



{\it Diagnostics--}
\begin{figure}[htbp]
	\centering
	\subfigure{\includegraphics[width=0.45\textwidth]{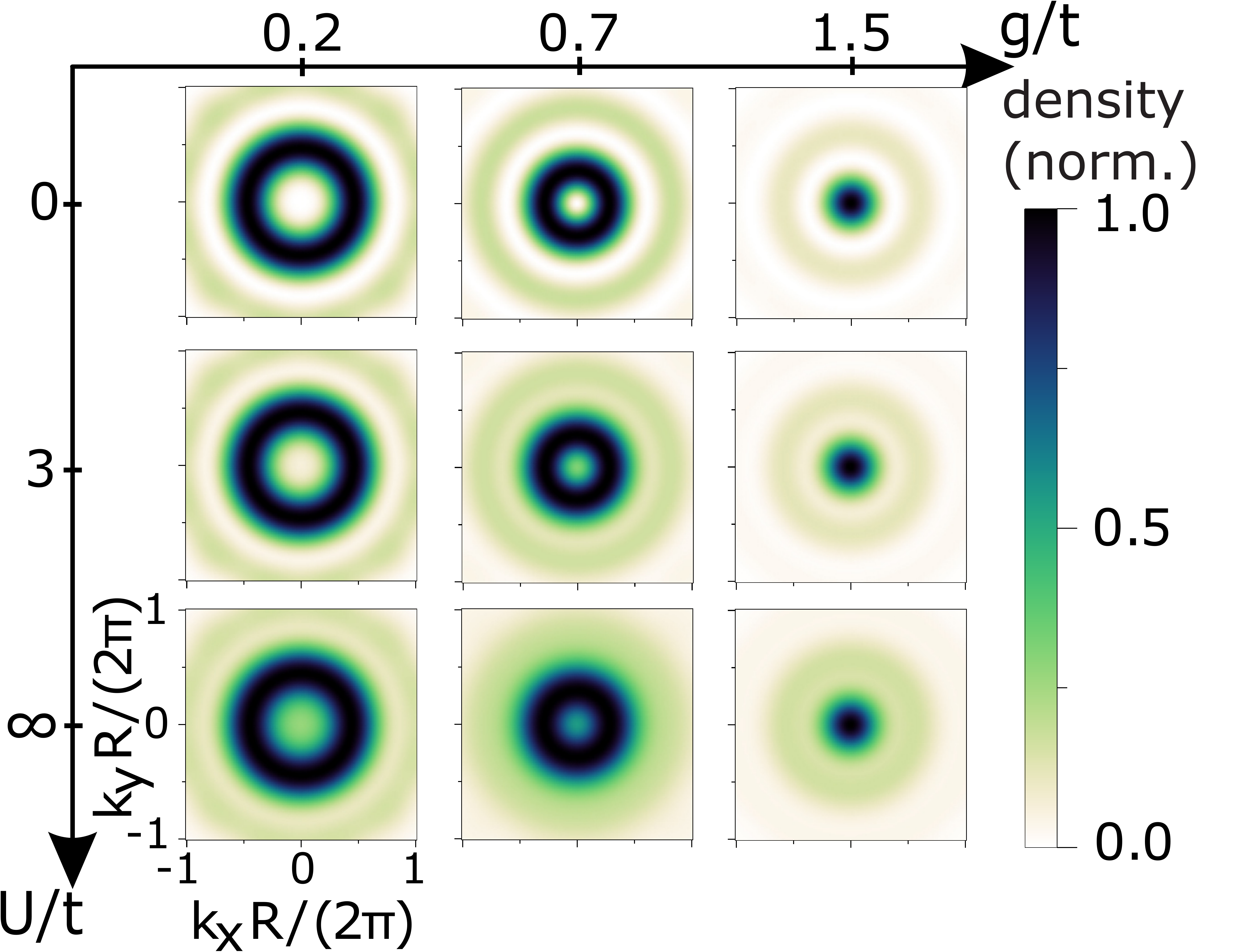}}
	\caption{Ground-state momentum distribution (time of flight) integrated along the z-axis (as defined in Fig.\ref{Sketch}a) for different values of $U/t$ and $g/t$ for two rings with 12 sites each, 6 bosons, ${\gamma=1}$, ${w=0}$ and total flux ${\phiA L=-\phiB L=3.12\pi}$. Details on the calculation are found in Appendix \ref{AppendixTOF} and \protect\cite{aghamalyan2015coherent}. 2D graphs show momentum $k$ in XY-plane in units of ring radius $R$. The inter-ring distance is $\frac{3}{5}R$. The phase winding and quasi-momentum is proportional to the diameter of the hole \revA{(area of low density at the center of the graphs)}.   By increasing $g/t$ from left to right, the phase winding and hole decreases to zero in two distinct steps. \revA{A hole visible in the centre indicates that the system is in the Vortex phase (left-most and central graphs). The disappearance of the hole and the appearance of a single peak at the centre indicates that the system is in the Meissner phase (right-most graphs).} With increasing interaction $U/t$, momentum distribution smears out. Density is normalized in each image. }
	\label{TOFXY}
\end{figure}
The flow of the ultracold atoms confined in ring-shaped optical potentials  can be read out  through  time-of-flight experiments\cite{amico2005quantum}.
We exploit such  a feature to tell apart Vortex and Meissner phases. 
Fig.\ref{TOFXY} shows the momentum distribution as a result of a time-of-flight experiment for different values of interaction and inter-ring coupling and ${\abs{\phi}L >\pi}$. 
The width of the annulus is proportional to the phase winding of the condensate, which only assumes integer multiples and is a measure for the condensate flow. \revA{The width of the hole is proportional to the $K_\text{c}$ parameter, as seen in Fig.\ref{CurrentChiralU}a.} From small to large coupling the number of phase windings changes from ${n=2}$ (${n=-2}$) in ring A (B) to ${n=0}$. In the Meissner phase, the phase winding is zero and we observe no hole in the time-of-flight.  We observe  that  no hole in the time of flight of the system in the Meissner phase arises for ${\abs{\phi}L<\pi}$ (flux smaller than a single flux quantum): {\it As a characteristic trait of the mesoscopic regime, the Vortex and Meissner phase can only be told apart above a threshold in $\phi$}.
Interaction smears out the distribution, but the hole is still well visible. Thus, the time of flight experiments are a reliable way to determine the phase winding for two coupled rings. 
\revA{To gain the momentum distribution of an individual ring, the population of the other ring could be destroyed (e.g. by exciting it with a laser), before releasing the remaining ring to be measured.}
Observing the condensate from different angles or tracing the time-evolution of the expansion could realize a way to gain more information about the quasi-momentum distribution. Co-expanding the rings with an additional condensate as phase reference \cite{eckel2014interferometric} can reveal the sign and superposition states of the phase winding.

{\em Conclusion--} 
\revA{We studied the currents and the different physical regimes that can be established in a Bose condensate confined in a ring ladder of  mesoscopic scale.  Such a system defines a  specific   atomtronic circuit with enhanced control\cite{Amico_Atomtronics,amico2014superfluid,li2008real}. Because of the closed geometry and  the mesoscale of the system,  the physics of the system can be related to notions  like quantum phase slips and persistent currents, that cannot be even defined in the previous studies concerning the open ladders. Specifically:
{\it i)} By tuning the inter-ring distance, mesoscopic currents along the legs are responsible for   an offset and quantized steps of the chiral current. With interaction, the smoothening of the steps reflects superpositions of different phase windings and quantum phase slips. 
{\it ii)} The ring-twist $w$ allows to vary continuously between rectangular  and triangular ladder configurations (see Fig.\ref{Sketch}a,b). As function of $w$, we found  a reentrant behaviour in the Meissner-Vortex phase diagram and in the chiral momentum (see Fig.\ref{CritFlux}). We note that  the dispersion is flat at the reentrance for ${\gamma=1}$ and ${\phi=\pi}$. This feature could be exploited to  simulate Weyl semi-metals\cite{nie2017scaling}.
{\it iii)}  Finally, we demonstrated with our approach that time-of-flight measurements  (in the plane of the rings) provide a feasible way to expose the physics of the mesoscopic system implied in the current and phase winding of the condensate (see Fig.\ref{TOFXY}). }

\revA{Recently, it has been shown that the current can reverse under specific conditions\cite{greschner2015spontaneous}. This poses an interesting subject to study with our set-up. 
The transition of the ring ladder between mesoscopic and macroscopic behaviour with interaction could be studied using DMRG.}
In momentum space, the ring-twist realizes an inter-ring coupling which depends on the quasi-momentum $k$ with ${g_k\propto1-w+w\expU{ik}}$. For an atom with fixed $k$, we can identify each ring as an internal state of pseudo-spins \cite{li2016spin}. This concept has been used to realize a supersolid\cite{li2017stripe}. With the twisted ring configuration, it could realize a nonlinear spin-orbit coupling to study new quantum phases. 

In possible future work a ladder with three or more legs (increasing the number of rings) can be considered  to study quantized edge currents analogue to the quantum Hall effect \cite{stuhl2015visualizing,mancini2015observation}. The entanglement inherent to interacting bosons could be used to create a protocol for quantum enhanced rotation sensing \cite{ragole2016interacting}. 

\begin{acknowledgments}
We thank A. Minguzzi, A. Leggett, W. Li, N. Victorin and T. Vekua for enlightening discussions. The Grenoble LANEF framework (ANR-10-LABX-51-01) is acknowledged for its support with mutualized infrastructure. We thank National Research Foundation Singapore and the Ministry of Education Singapore Academic Research Fund Tier 2 (Grant No. MOE2015-T2-1-101) for support.
\end{acknowledgments}


\bibliography{library}
\appendix
\section{Ring-twist by introducing diagonal tunneling}\label{AppendixDiag}
\begin{figure}[htbp]
	\centering
	\subfigimg[width=0.24\textwidth]{a}{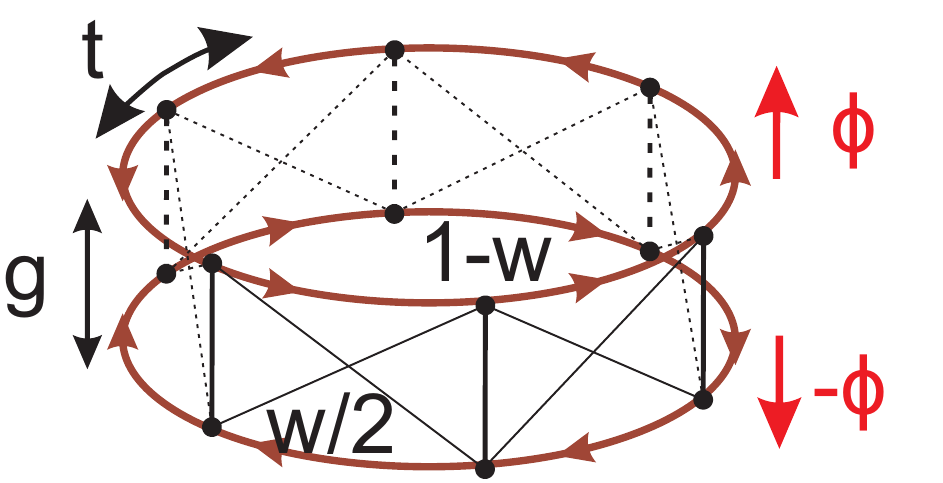}\hfill
	\subfigimg[width=0.24\textwidth]{b}{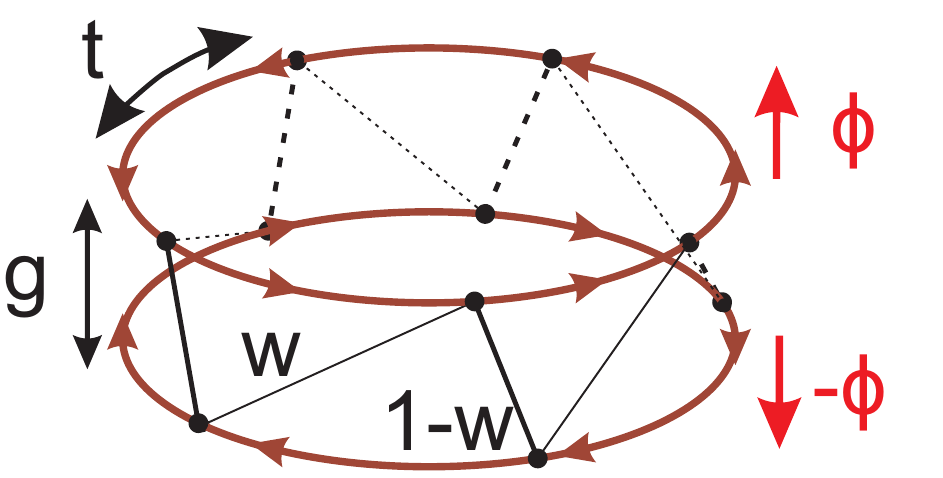}
	\caption{Schematics of the two rings with ring-twist $w$ for \idg{a} diagonal configuration ($\gamma=0$) and \idg{b} physical twist ${\gamma=1}$. For \idg{a}, diagonal coupling with strength $w/2$ is introduced. For \idg{b}, lattice sites on ring A are shifted relative to ring B, such that a site on ring A couples to two other sites on ring B with rate $w$ and ${1-w}$.}
	\label{FigureSketchTwist}
\end{figure}
\begin{figure}[htbp]
	\centering
	\includegraphics[width=0.4\textwidth]{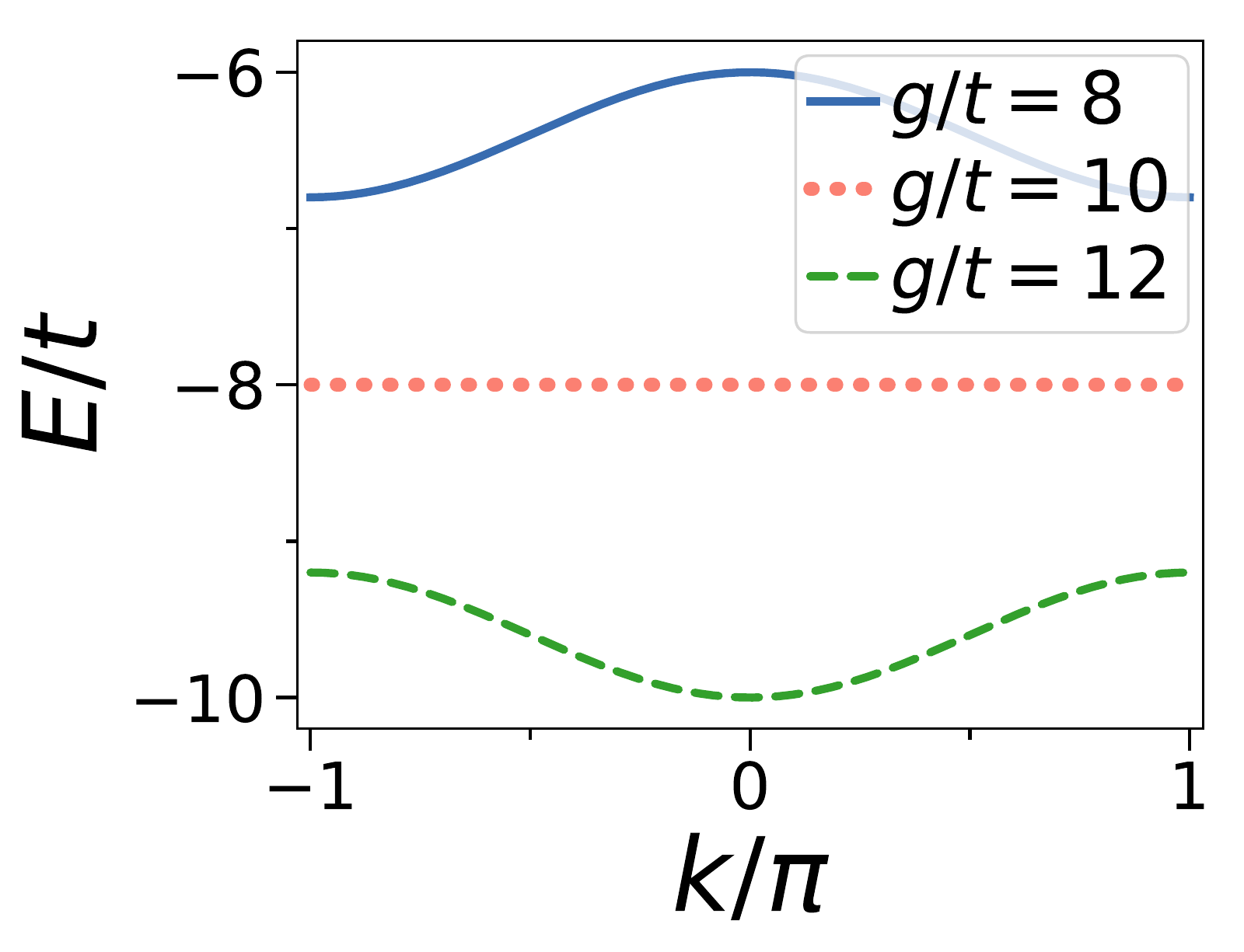}
	\caption{Dispersion relation for diagonal twist configuration with ${w=0.2}$ and ${\phi=\pi}$. Dispersion is flat for ${g=\frac{2t}{w}=10}$.}
	\label{DispersionTwist}
\end{figure}

For the ladder with ring-twist two different Hamiltonians are proposed. In this section, the ring-twist is engineered by a diagonal inter-ring tunneling with ${\gamma=0}$ as defined in the main text. This corresponds to a strong next-nearest-neighbor inter-ring tunneling. The setup is sketched in Fig.\ref{FigureSketchTwist}a.
To create the lattice in the lab, we propose the following procedure. First, a two ring lattice is created by a blue-detuned laser. The ring sites of each lattice are  vertically aligned, which has been successfully demonstrated in the lab \cite{amico2014superfluid}. Next, a second tightly focused red-detuned single ring lattice is projected in between the two rings. This lattice is chosen such that it modifies the tunneling rate between the two blue-detuned lattice rings. By modifying this second potential, nearest-neighbor and next-nearest neighbor inter-tunneling rates between the two lattice rings can be modified. This red-detuned lattice is used to reduce the nearest-neighbor inter-ring tunneling, and increase next-nearest neighbor tunneling. Then, we define the ring-twist as the difference of nearest and next-nearest neighbor inter-ring tunneling contributions. The new ring-ring interaction Hamiltonian is then
\begin{align*}
\mathcal{H_\text{I}}={}&\sum_{m=1}^{L}-g\left[(1-w)\cn{a}{m}\an{b}{m}+\frac{w}{2}\cn{a}{m}\an{b}{m+1}+\frac{w}{2}\cn{a}{m}\an{b}{m-1}\right]+\text{h.c.} \; .
\end{align*}
In this Hamiltonian there are now three inter-ring tunneling contributions. Our calculations show that the same re-entrant behaviour is found in nearly the same parameter space as in the ${\gamma=1}$ configuration.
The dispersion relation (derivation for case without ring-twist found here\cite{kelecs2015mott}) is
\begin{align*}
E_\pm=&{}-2t\cos(k)\cos(\phi)\\
&\pm\sqrt{[g(1-w(1-\cos(k)))]^2+(2t\sin(k)\sin(\phi))^2}\;.
\end{align*}
The negative branch is flat and independent of quasi-momentum $k$ for ${g=\frac{2t}{w}}$ and ${\phi=\pi}$ when $0<w<0.5$. The dispersion relation in that region is plotted in Fig.\ref{DispersionTwist}. This corresponds to massless particles and can be used to simulate Weyl semi-metals \cite{nie2017scaling}.
\revA{The Meissner phase is characterized by a single minimum of quasi-momentum $k$ of the dispersion relation at ${k=0}$ or ${k=\pi}$. The Vortex phase has a two degenerate minima at $\pm k_\text{min}$ (${t=1}$)
\begin{align*}
&k_\text{min}/\pi=\arccos\left[\left(g^2(1-w)w(4-g^2 w^2)\right.\right.\\
&\left.\pm2\sqrt{(-4+g^2 w^2)(-2-g^2(1-2w)+2\cos(2\phi))\sin(2\phi)^2}\right)\\
&\left./((-4+g^2 w^2)(g^2 w^2-4\sin(\phi)^2))\right]
\end{align*}
The positive branch is for ${\phi<0.5\pi}$, the negative for ${\phi>0.5\pi}$. Whenever the above equation is real-valued, the system is in the Vortex phase. The occupation number is $n_\text{A}(k)=N_\text{p}/2\cos^2(\theta_k)$, $n_\text{B}(k)=N_\text{p}/2\sin^2(\theta_k)$, with $\theta_k=\frac{1}{2}\arctan\left(\frac{g(1-w(1-\cos(k)))}{2t\sin(k)\sin(\phi)}\right)$. For a system with infinite number of sites, chiral momentum and current can be calculated by plugging the value for $k_\text{min}$, $n_\text{A}(k)$ and $n_\text{B}(k)$ into Eq.\ref{Eqchiralmomentum} and Eq.\ref{Eqchiralcurrent} respectively.}
The phase diagram can be calculated from the minima of the dispersion relation. The lower transition line is given by
\begin{align*}
\cos{\left(\phi_\text{c}\right)}={}&-\frac{g}{4t}+\sqrt{1+\left(\frac{g}{4t}\right)^2(1-4w)^2}
\end{align*}
and the upper transition line
\begin{align*}
g_\text{c}/t={}&\frac{1}{(1-2w)w}\left[1-6w+8w^2+(1-2w)\cos(2\phi)\pm\right.\\
&\left.\sqrt{2}\sqrt{(1-2w)^3\cos(\phi)^2[1-6w+(1+2w)\cos(2\phi)]}\right]^{\frac{1}{2}}~,
\end{align*}
for solutions with imaginary part zero.

\begin{figure}[htbp]
	\centering
	\subfigimg[width=0.24\textwidth]{a}{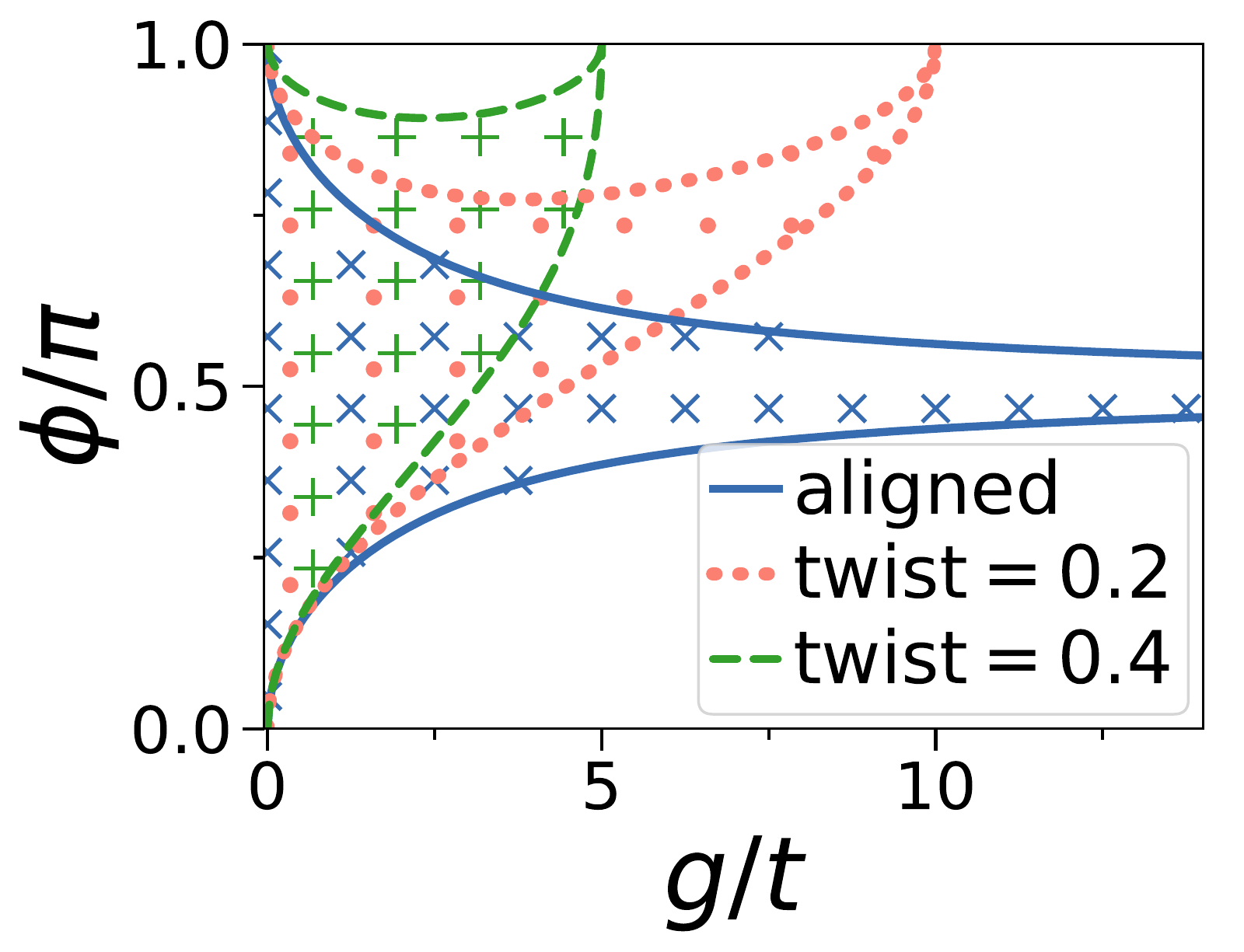}\hfill
	\subfigimg[width=0.24\textwidth]{b}{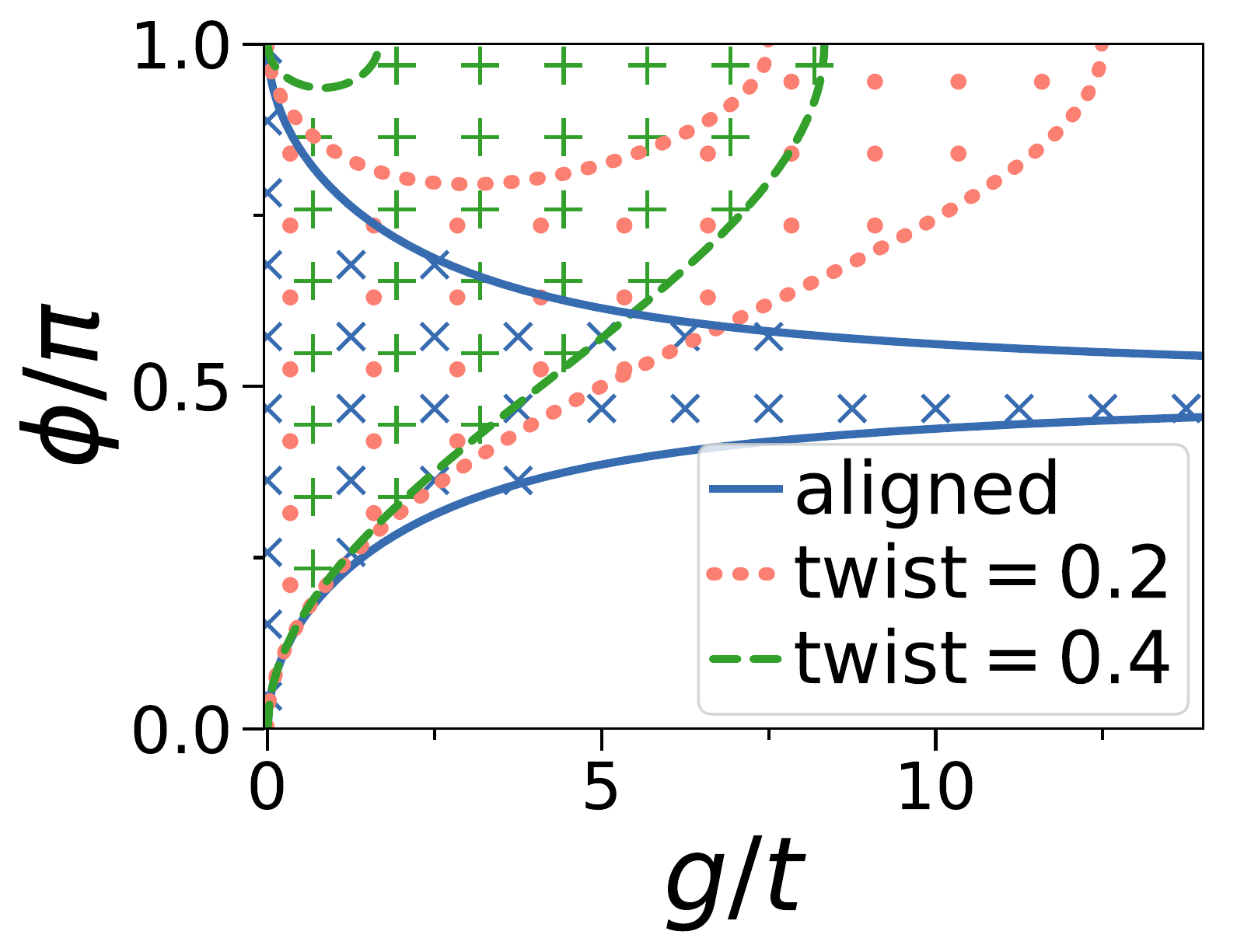}
	\caption{Phase diagram for zero interaction with inter-ring tunneling $g$ plotted against flux per site $\phi$ for diagonal tunneling ${\gamma=0}$. The Vortex phase is denoted by the dotted/crossed area. {\bfseries a)} shows phases for three nearest-neighbor inter-ring tunneling contributions in a diagonal configuration. At $\phi=\pi$ and $\frac{g}{t}=\frac{2}{w}$, where Vortex and Meissner phase merge, the dispersion relation is constant.	{\bfseries b)} shows phases for two rings in the ${\gamma=1}$ configuration. Both graphs depend strongly on ring-twist for ${\phi>0.5\pi}$. They show a reentrant behaviour between the phases along the $g/t$ axis for intermediate ring-twist and ${\phi\approx0.8\pi}$. 
	}
	\label{FigureVortex}
\end{figure}

\begin{figure}[htbp]
	\centering
	\subfigimg[width=0.24\textwidth]{a}{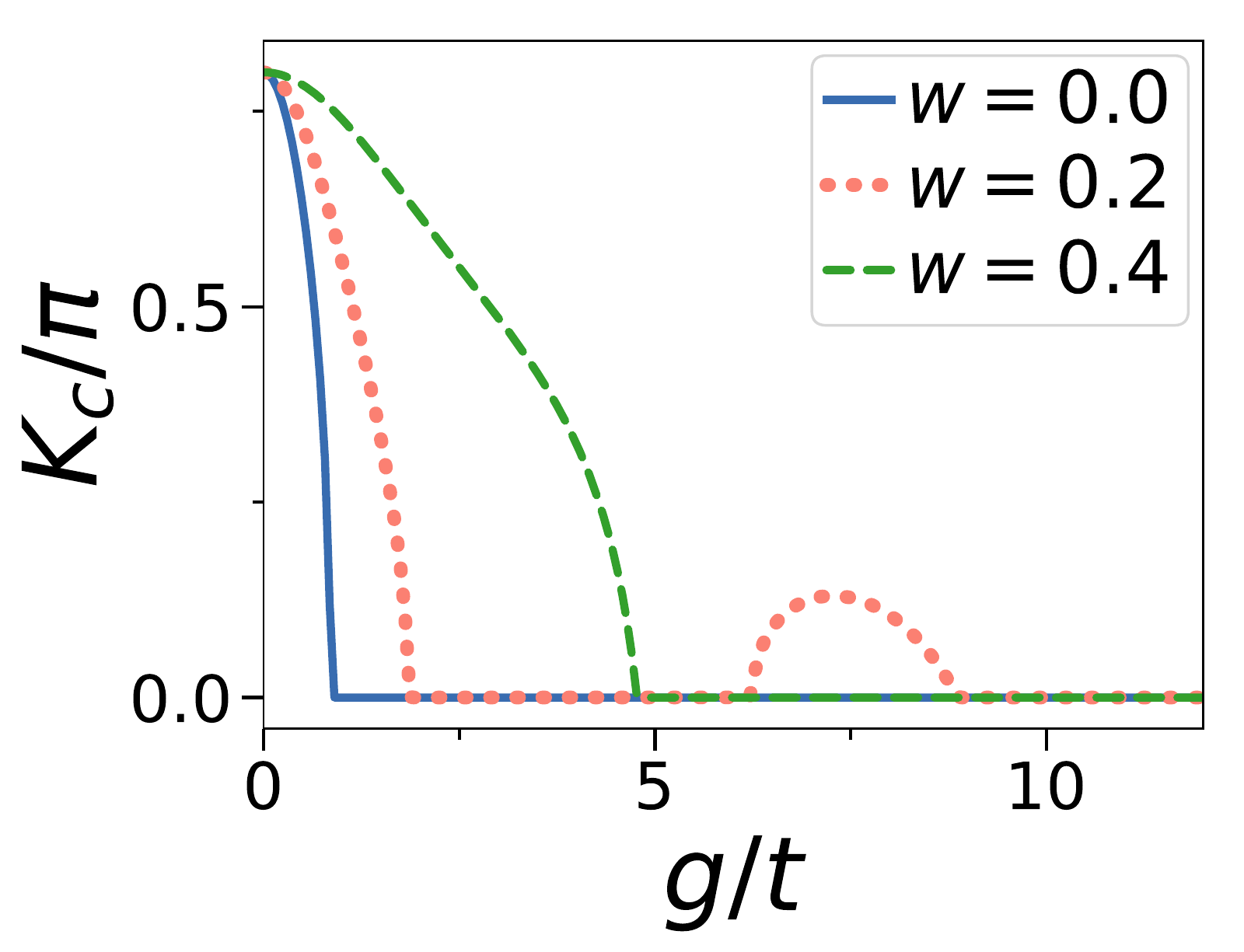}\hfill
	\subfigimg[width=0.24\textwidth]{b}{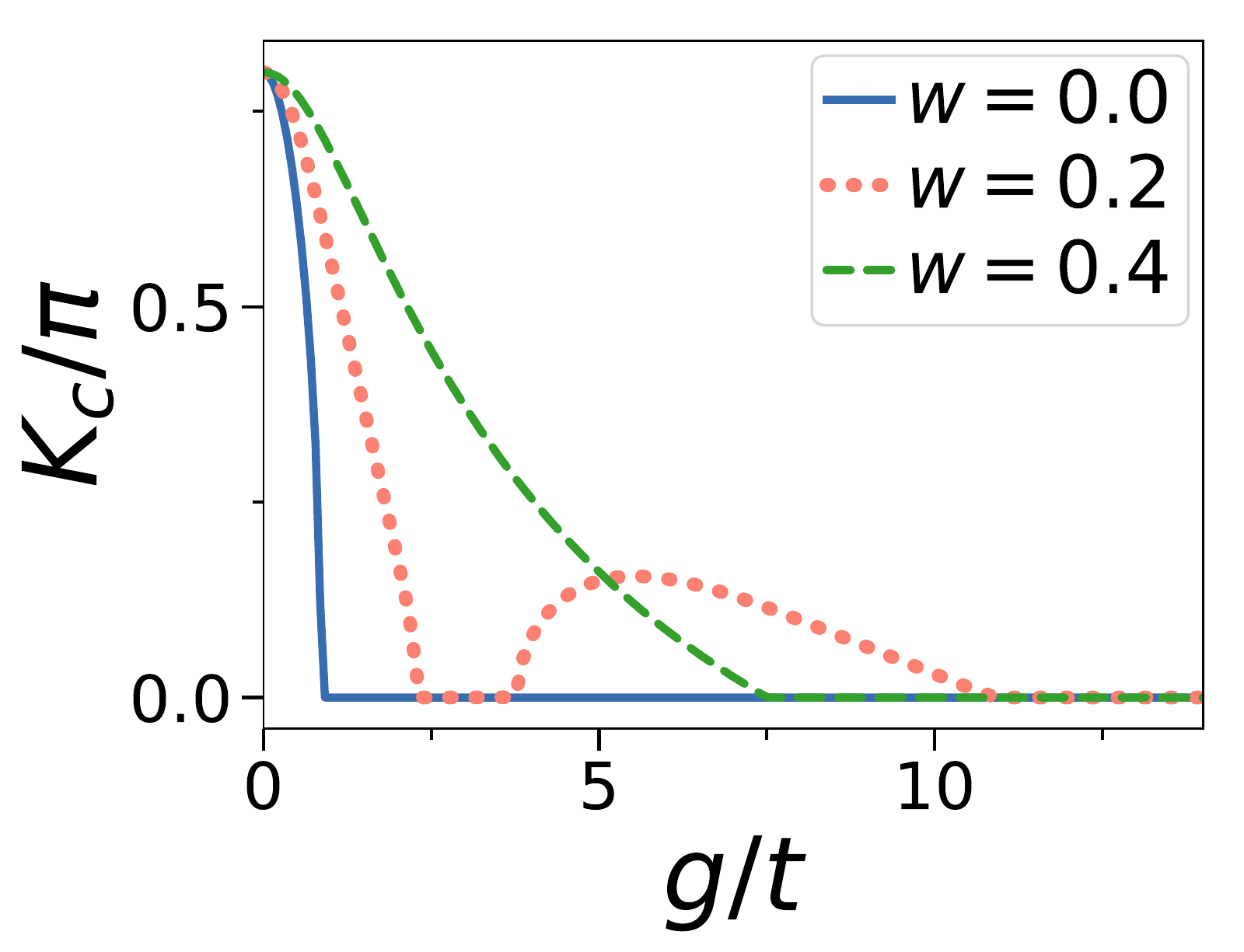}
	\caption{Chiral momentum against inter-ring coupling $g$ for different values of ring-twist $w$. {\bfseries a)}  For diagonal inter-ring tunneling (${\gamma=0}$). 
	{\bfseries b)} For the shifted configuration (${\gamma=1}$).}
	\label{ChiralCurrent}
\end{figure}

\begin{figure}[htbp]
	\centering
	\subfigimg[width=0.24\textwidth]{a}{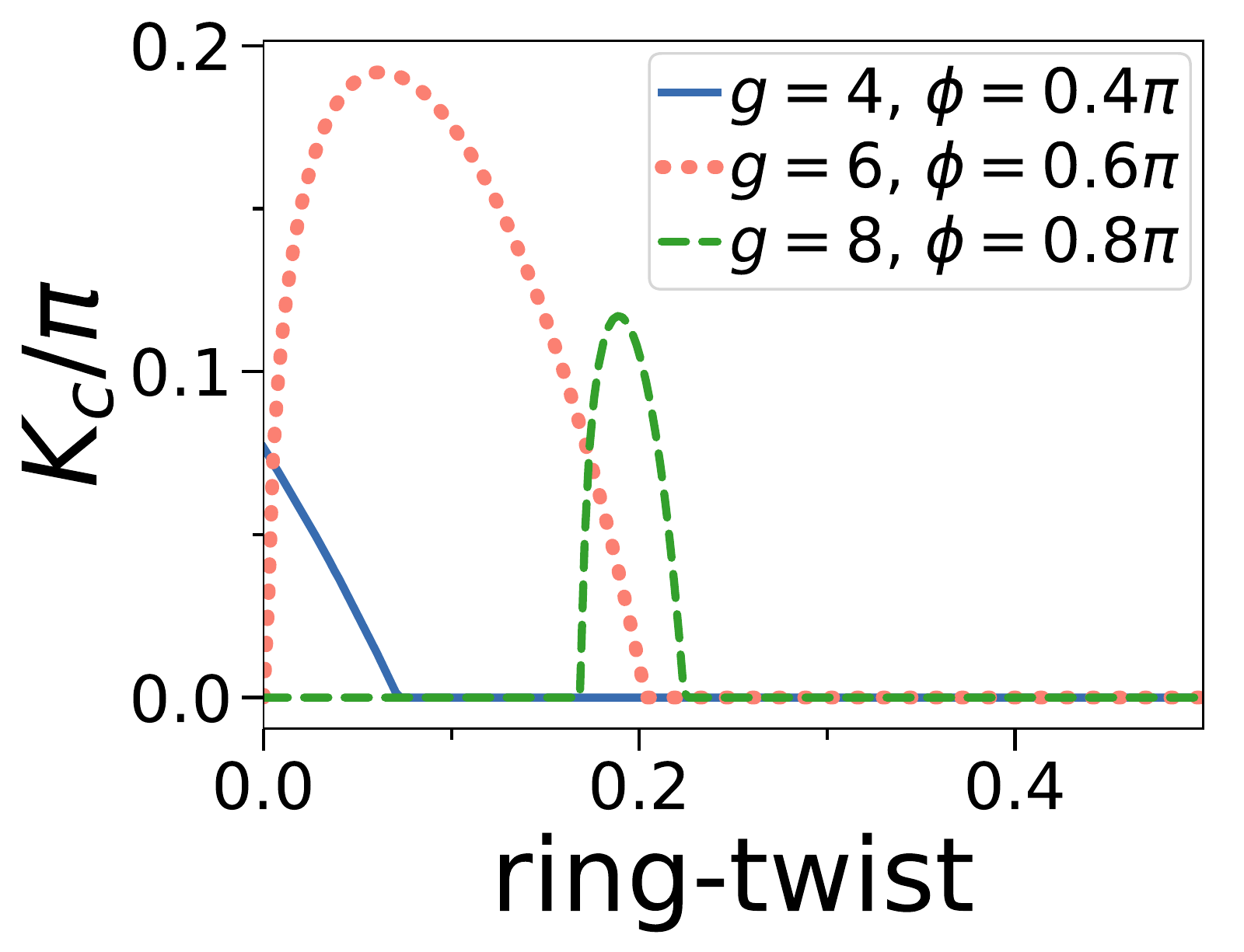}\hfill
	\subfigimg[width=0.24\textwidth]{b}{chiralKTwistN1DTwistN10000.pdf}
	\caption{Chiral momentum against ring-twist $w$ for different parameters. {\bfseries a)} For diagonal inter-ring tunneling (${\gamma=0}$). 
		{\bfseries b)} For the shifted configuration (${\gamma=1}$).}
	\label{ChiralCurrentTwist}
\end{figure}

The resulting phase diagram is plotted in Fig.\ref{FigureVortex}a. Chiral momentum against inter-ring coupling $g/t$ is plotted in Fig. \ref{ChiralCurrent}a and against ring-twist $w$ in Fig. \ref{ChiralCurrentTwist}a.

\section{Ring-twist created by relative shift of the rings}\label{AppendixRotation}
We discuss the realization of the twist $w$ by introducing a relative shift of one ring lattice to the other, which corresponds to ${\gamma=1}$. A cartoon of the configuration is shown in Fig.\ref{FigureSketchTwist}b. The Hamiltonian is 
\begin{align*}
\mathcal{H_\text{I}}={}&\sum_{m=1}^{L}-g\left[(1-w)\cn{a}{m}\an{b}{m}+w\cn{a}{m}\an{b}{m+1}\right]+\text{h.c.} \; ,
\end{align*}
and the dispersion relation
\begin{align*}
E_\pm=&{}-2t\cos(k)\cos(\phi)\\
&\pm\sqrt{g^2(1-2w(1-w))(1-\cos(k))+(2t\sin(k)\sin(\phi))^2}\;.
\end{align*}
\revA{The Meissner phase is characterized by a single minimum of quasi-momentum $k$ of the dispersion relation at ${k=0}$ or ${k=\pi}$. The Vortex phase has a two degenerate minima at $\pm k$ (${t=1}$)
\begin{align*}
k/\pi=&\arccos\left[\frac{1}{4\sin(\phi)^2}(g^2(1-w)w\pm\right.\\
&\left.\sqrt{\cos(\phi)^2(g^2(1-w)^2+4\sin(\phi)^2)(g^2w^2+4\sin(\phi)^2)}\right]\;.
\end{align*}
The positive branch is for ${\phi<0.5\pi}$, the negative for ${\phi>0.5\pi}$. Whenever the above equation is real-valued, the system is in the Vortex phase. The occupation number is $n_\text{A}(k)=N_\text{p}/2\cos^2(\theta_k)$, $n_\text{B}(k)=N_\text{p}/2\sin^2(\theta_k)$, with $\theta_k=\frac{1}{2}\arctan\left(\frac{g\sqrt{1-2w(1-w)(1-\cos(k))}}{2t\sin(k)\sin(\phi)}\right)$. For a system with infinite number of sites, chiral momentum and current can be calculated by plugging the value for $k_\text{min}$, $n_\text{A}(k)$ and $n_\text{B}(k)$ into Eq.\ref{Eqchiralmomentum} and Eq.\ref{Eqchiralcurrent} respectively.}
The transition between Meissner and Vortex phase can be calculated from the minima of the dispersion relation, and is given by
\begin{align*}
\cos{\left(\phi_\text{c}\right)}={}&-\frac{g}{4t}+\sqrt{1+\left(\frac{g}{4t}\right)^2(1-2w)^2}~,\\
\cos{\left(\phi_\text{c}\right)}={}&\frac{g}{4t}(1-2w)-\sqrt{1+\left(\frac{g}{4t}\right)^2}~.
\end{align*}
The phase diagram is plotted in Fig.\ref{FigureVortex}b. Chiral momentum against inter-ring coupling $g/t$ is plotted in Fig.\ref{ChiralCurrent}b and against ring-twist $w$ in Fig.\ref{ChiralCurrentTwist}b.

Both rectangular  and triangular ladder configurations ($w=0$ and $w=0.5$) may be realized by Laguerre-Gauss beams as outlined in \cite{aghamalyan2013effective}.

We propose different options to create a generic ring-twist $w$:
First, spatial light modulators (SLM) or digital micromirror devices (DMD) can create arbitrary potentials by shaping wavefronts. By using two devices and two red detuned light fields with different wavelength to circumvent interference one can envision the following procedure.
Beam propagation is along $z$-direction. In a plane orthogonal to $z$, two ring lattices at positions $z_\text{a}$ and $z_\text{b}$ are created. Each ring $\alpha$ is created by one of the SLMs with wavelengths $\lambda_\alpha$ by tightly focusing the light at position $z_\alpha$. The light pattern created by the SLM is programed such that the lattice sites of ring A are shifted relative to ring B as depicted in the cartoon Fig.\ref{FigureSketchTwist}.

However, the on focused light intensity of the image at $z_\text{b}$ will influence the potential the atoms see in the plane of $z_\text{a}$ as the two focal points are close together. One can now change the potential at $z_\text{a}$ with the SLM, until a ring lattice shaped potential with the additional stray light from the beam focused at $z_b$ is restored. The change of the potential at $z_\text{a}$ will of course influence the total intensity at $z_\text{b}$. Now, the same procedure has to be done for the plane $z_\text{b}$. This will be an iterative process for both focal points until a reasonable two ring pattern is formed. This procedure has to be pre-calculated for each value of the ring-twist. However as of now, no such experimental realization exists and the general convergence of this special iterative process has to be proven.

Alternatively, the ring-twist could be created with holographic methods\cite{latychevskaia2016inverted}. It has been shown that using an SLM spiral pattern in direction of beam propagation can be created. The ring-twist of the two rings corresponds to a relative shift of lattice sites in ring A and B. Using the same holographic method and a red-detuned laser, a ring lattice tube could be arranged with a spiral pattern in z-direction. By imprinting a intensity modulation in direction of beam propagation (e.g. by focusing a blue-detuned laser in the plane orthogonal to propagation), separated rings are formed. The spiral winding realizes the ring-twist.

Finally, the two rings could be created concentrically in the same plane, with two different radii. This way, creating and addressing the rings becomes simpler as only a single SLM or DMD is required to create any potential shape in 2 dimensions. However, as the two rings have different radii, an asymmetry in the two rings are introduced in the intra-ring hopping. This asymmetry could be corrected by modifying each ring potential separately

\begin{figure}[htbp]
	\centering
	\subfigimg[width=0.24\textwidth]{a}{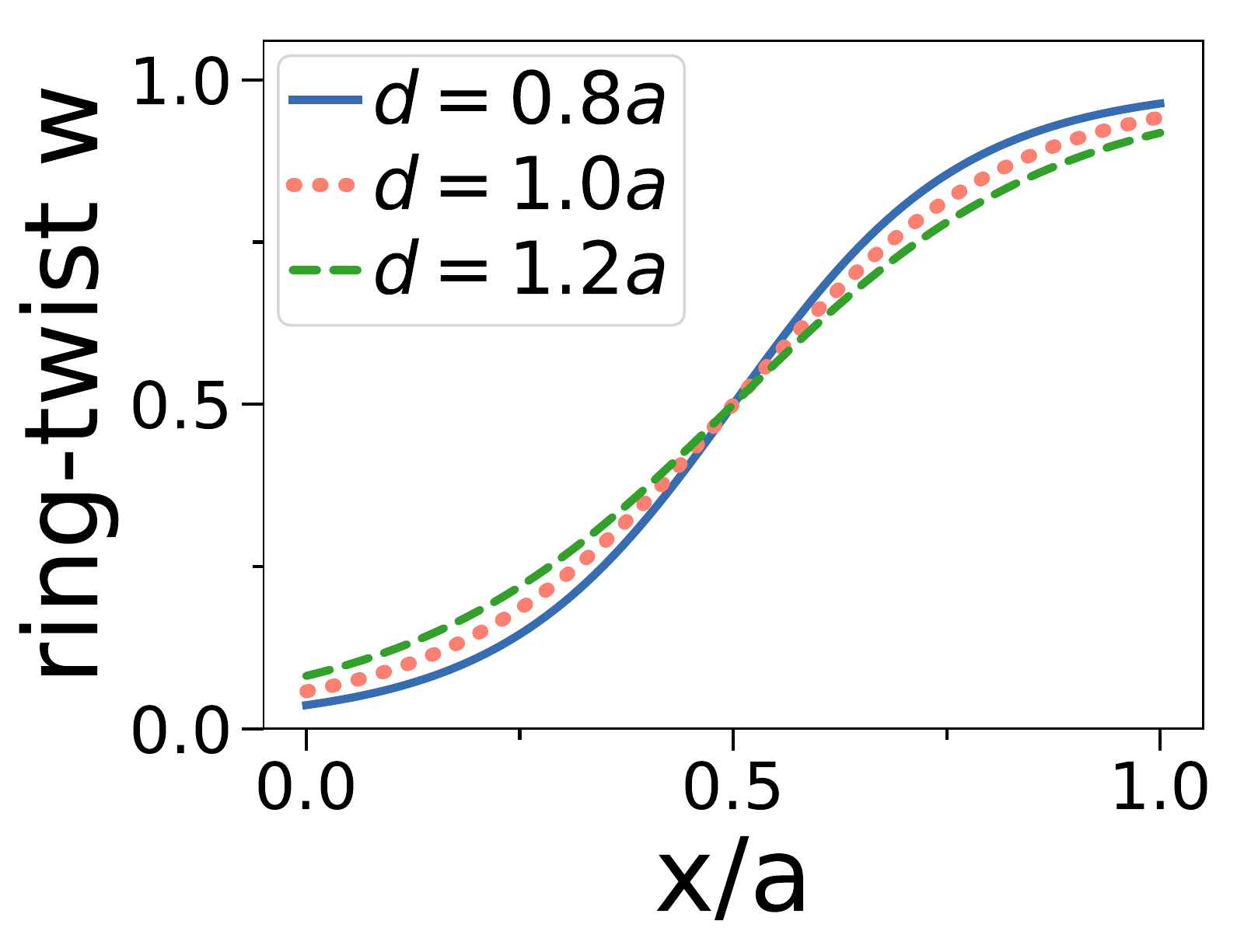}\hfill
	\subfigimg[width=0.24\textwidth]{b}{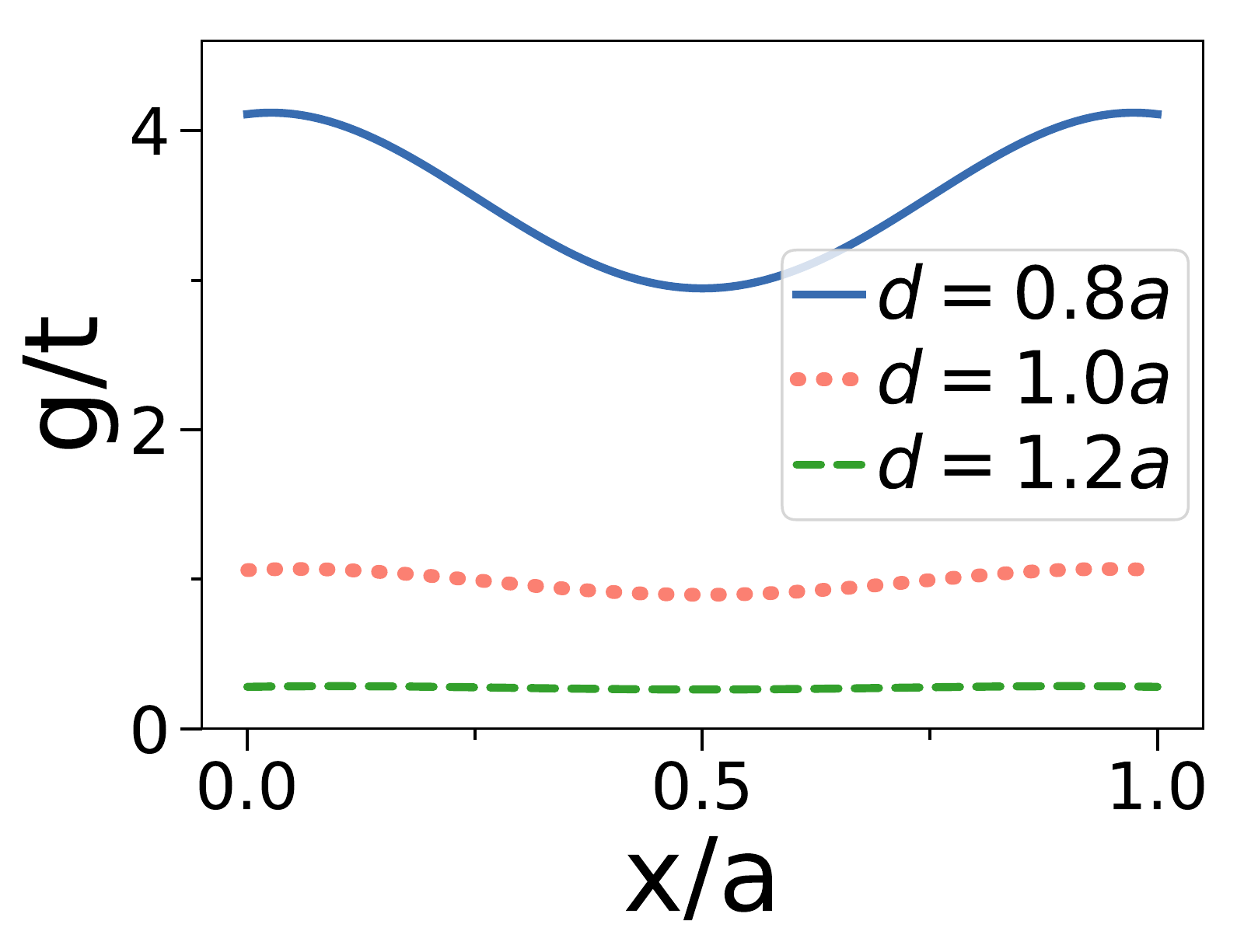}
	\caption{{\bfseries a)} ring-twist $w$ as defined in the paper plotted against the relative shift $x/a$ of ring A and B in units of lattice constant $a$. 	
		{\bfseries b)} ratio inter-ring coupling to intra-ring coupling $g/t$ plotted against the relative shift $x/a$ of ring A and B in units of lattice constant $a$.}
	\label{FigureTunneling}
\end{figure}

In the discussion so far the ring-twist parameter $w$ is a linear parameter in the lattice Hamiltonian. We investigate how this linear parameters relates to an actual twisted two lattice ring configuration.
The tunneling can be calculated with the WKB approximation \cite{aghamalyan2013effective}
\begin{equation*}
g=4\sqrt{\frac{\hbar}{\sqrt{2m}}}\frac{V_0^{3/4}}{\sqrt{s}}\expU{-\frac{\sqrt{2mV_0}}{\pi \hbar}s}~,
\end{equation*}
where $s$ is the distance between two sites.
We plot the tunneling rates with that function and our setup in the attached Figure \ref{FigureTunneling}. $a$ is the distance between the lattice sites of a ring, $g/t$ the ratio of inter-ring tunneling and intra-ring tunneling, $d$ the inter-ring distance, $x/a$ the relative shift in space of lattice sites between ring A and B. $x/a=0$ or $x/a=1$ corresponds to both rings aligned, $x/a=0.5$ corresponds to a triangular configuration. 
We assume Rubidium 87 atoms and rings with $L=12$ sites, radius $R=8\mu\text{m}$. A comparable lattice configuration has been realized in \cite{amico2014superfluid}. We assume a potential barrier of $V=20E_\text{recoil}$.

As seen in the left figure, the linear parameter ring-twist $w$ changes non-linearly with the physical shift of the rings. However, the change occurs in a continuous and monotonous fashion and covers nearly all possible values of ring-twist. By knowing this function, the physical shift of the rings can be related to our linear parameter $w$.

In the right graph, the ratio $g/t$ is plotted against the physical ring shift $x/a$. We notice that the inter-ring tunneling rate changes by a small factor. This means, that while physical twisting the rings, the inter-ring tunneling changes as well by a small amount. There are two ways to address this: Either the change in tunneling rate has to be accounted for in the evaluation of the experiment, or the inter-ring distance has to be tuned accordingly to keep the tunneling rate constant.

\section{Mesoscopic currents}\label{chiralcurrentApp}
\begin{figure}[htbp]
	\centering
	\subfigure{\includegraphics[width=0.4\textwidth]{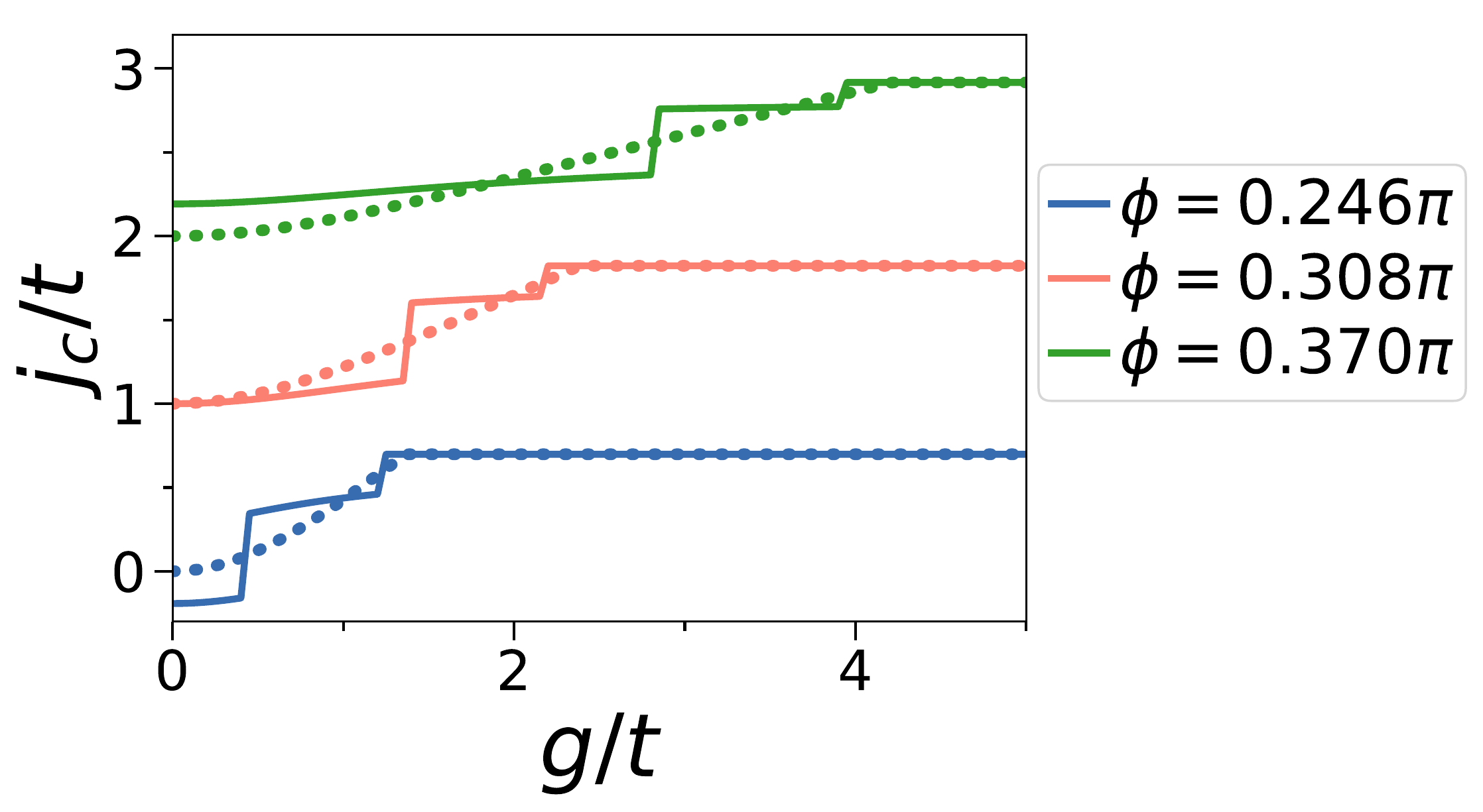}}
	\caption{Chiral current $j_\text{c}$ against inter-ring coupling $g/t$ to compare current for mesoscopic (${L=13}$, solid) and macroscopic (${L=\infty}$, dotted) number of sites. The chiral current for rings with a large number of sites (dotted curves) start at zero current and increase (Vortex), until they reach a constant plateau (Meissner). 
	For mesoscopic ring sizes (solid line) the persistent current creates an offset relative to zero in the current for small inter-ring couplings. The offset scales vanishes with increasing number of sites $L$, and the sign and value depend on the total flux in the rings. The curves follow the macroscopic case in a step-wise fashion, which is due to flux quantization in the rings.
	To enhance visibility, the curves for different flux are respectively shifted by 1.}
	\label{chiralcurrent}
\end{figure}

In Fig.\ref{chiralcurrent}, we study the chiral current (as defined in Eq.\ref{Eqchiralcurrent}) for mesoscopic and macroscopic number of sites. We find that for large number of sites the current is zero for small inter-ring couplings, while we observe a persistent current for a finite number of sites. The sign of the persistent chiral current changes with the total flux in the rings ${\Omega=\phi L}$.

\section{Dispersion relation}
\begin{figure}[htbp]
	\centering
	\subfigimg[width=0.24\textwidth]{a}{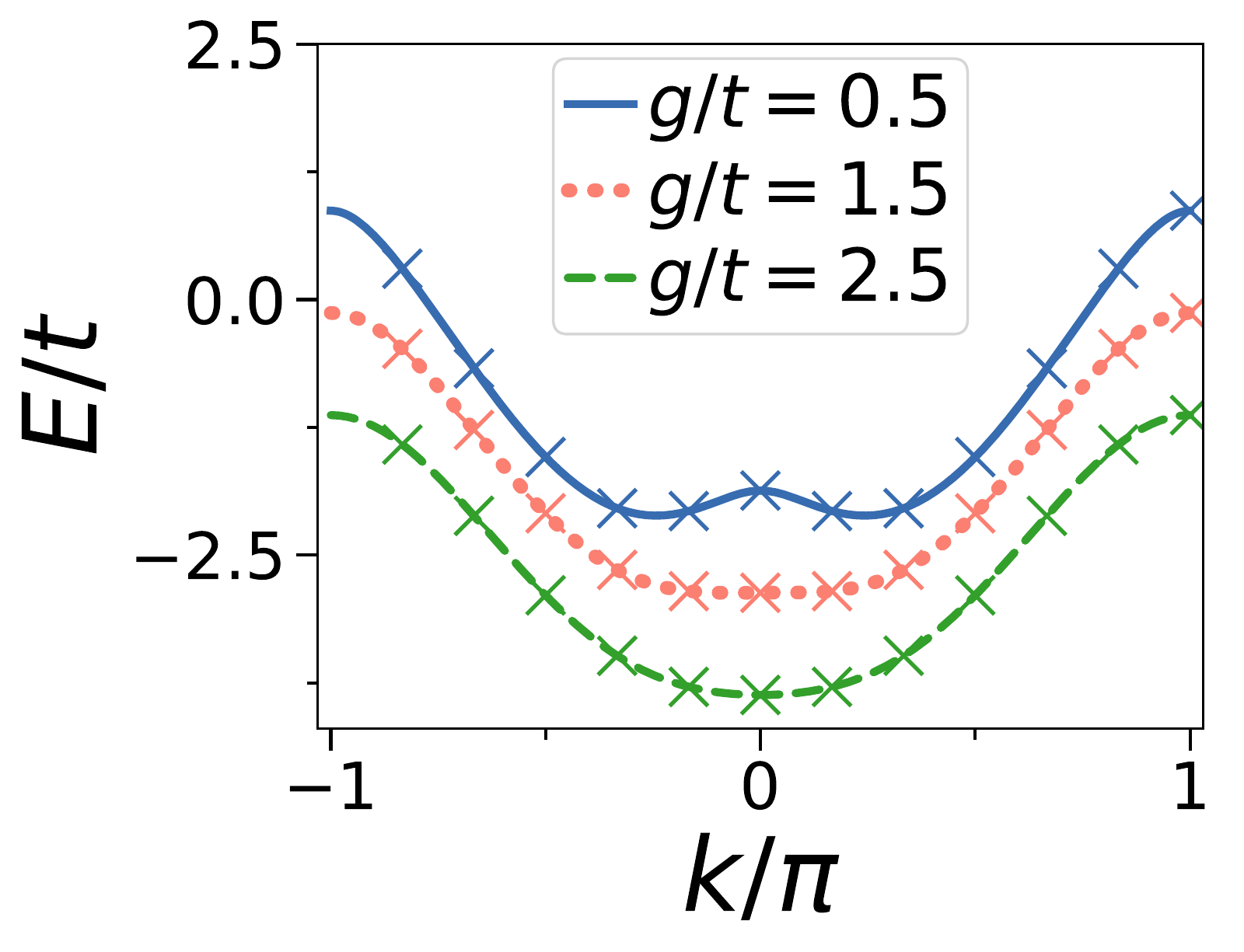}\hfill
	\subfigimg[width=0.24\textwidth]{b}{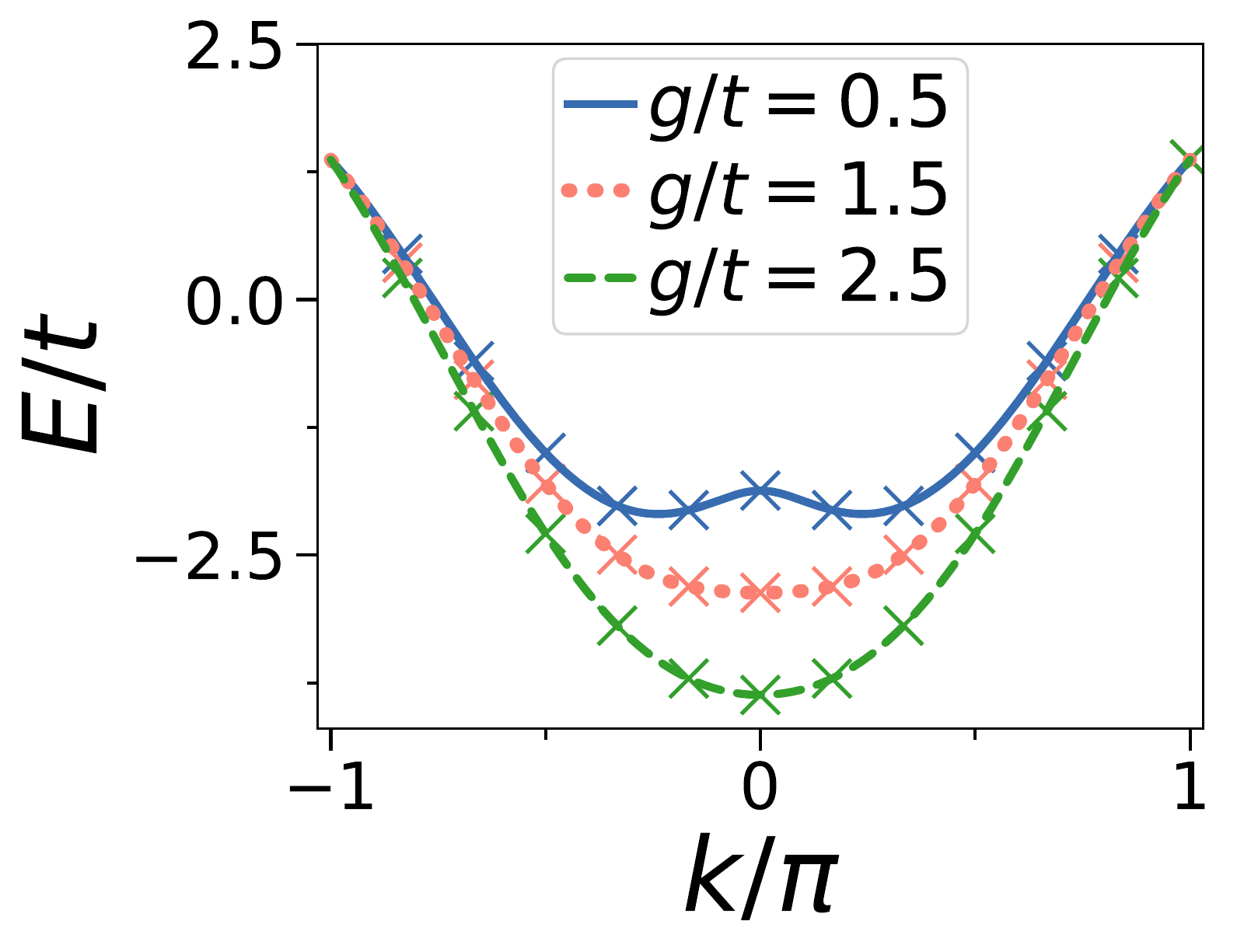}
	\caption{Dispersion relation (energy against quasi-momentum $k$) of the lower band for ${\gamma=1}$ with flux per site ${\phi_\text{A}=-\phi_\text{B}=0.26\pi}$. {\bfseries a)} shows aligned rings (${w=0}$) and {\bfseries b)} with maximal ring-twist ${w=0.5}$ for different values of inter-ring coupling $g/t$. Lines show infinitely sized system, crosses for two rings with 12 sites per ring where only $k$ satisfying the boundary condition ${k_n=2\pi n/L}$ are allowed. The dispersion relation has either a single minimum (Meissner) or two minima ${k=\pm k_0}$ (Vortex). For infinitely number of sites, the ground state $k$ changes continuously with the external parameter, whereas for finite size systems $k$ switches between its discrete values.}
	\label{SpectrumQuant}
\end{figure}

\begin{figure}[htbp]
	\centering
	\subfigure{\includegraphics[width=0.24\textwidth]{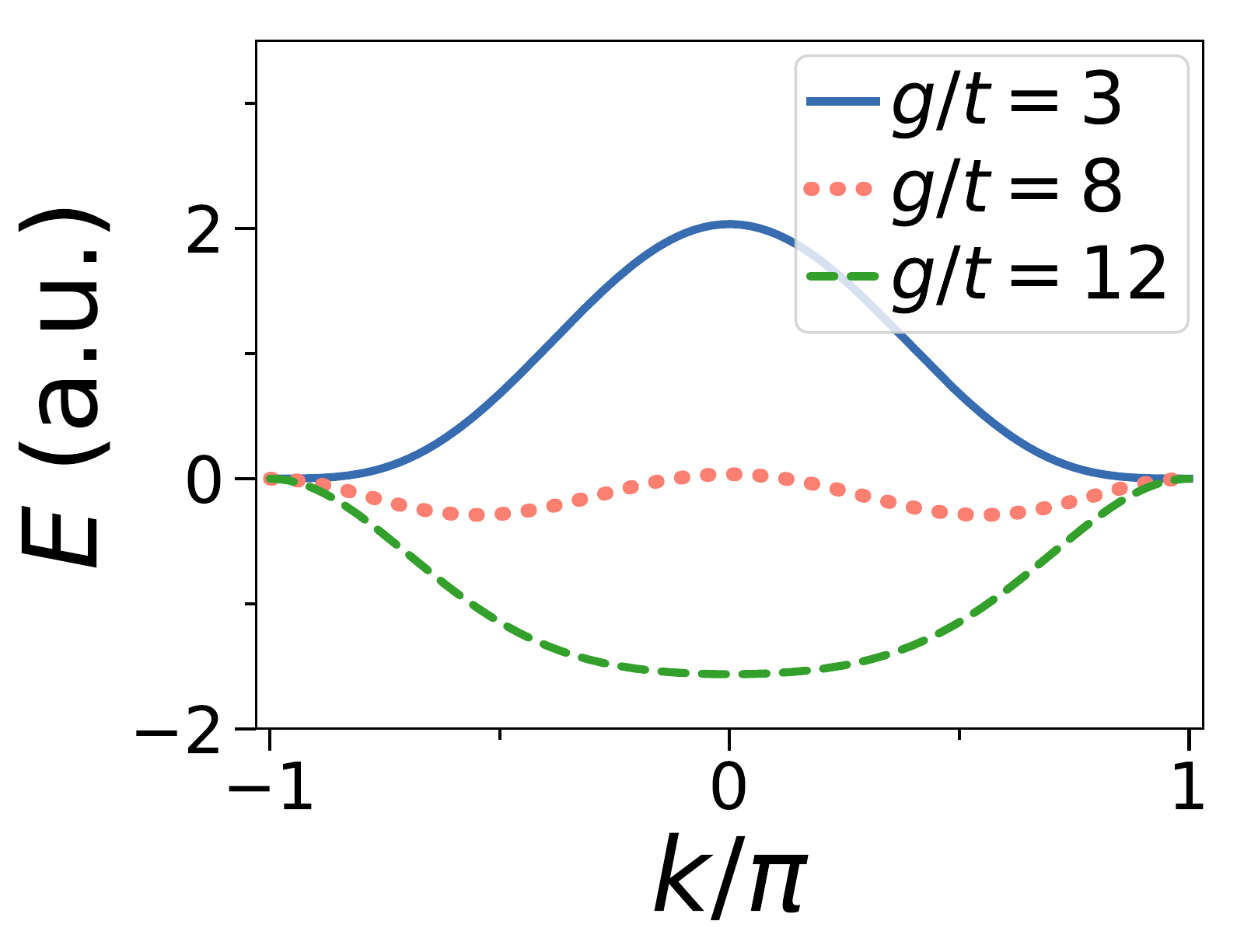}}
	\caption{Dispersion of the lower band for ring-twist ${w=0.2}$, ${\gamma=1}$, ${\phi=0.8\pi}$ for different values of inter-ring coupling $g/t$. For very small values of $g/t$, two minima exist (not shown). With increasing $g/t$, Meissner phase with one minimum at ${k=\pi}$ (solid blue), Vortex phase with two minima (dotted orange) and finally Meissner phase again with minimum at ${k=0}$ (dashed green).}
	\label{SpectrumTwist}
\end{figure}

Periodic boundary conditions quantize the quasi-momentum to $k_n=\frac{2\pi n}{L}$, with integer $n$. As an example, we show the allowed quasi-momenta for $L=12$ with cross symbols in Fig. \ref{SpectrumQuant} with ${\gamma=1}$ for zero and maximal ring-twist. 
The dispersion relation with intermediate ring-twist is plotted as full line in Fig.\ref{SpectrumTwist}. The Meissner state is characterized by a single minimum at quasi momentum ${k=0}$ or ${k=\pi}$ and the Vortex phase is characterized by two degenerate minima. 

\section{Qubit dynamics of the ring ladder}\label{qubitdynamics}
It has been shown, that for sufficiently large $L$, the dynamics of the rectangular ladder system ${w=0}$ is governed by an effective two-level system. Here, we study the configuration leading to the qubit dynamics. In the present case of moderate $L$, we find for ${\phiA=-\phiB=\pi/2}$
the energy gap is well defined for a wide parameter range. However, the first and second excited states results to be degenerate. This feature may render controlled addressing of only the two lowest levels as an effective qubit difficult. For this parameter regime, the system is always in a Vortex state.
\begin{figure}[htbp]
	\centering
	\subfigure{\includegraphics[width=0.4\textwidth]{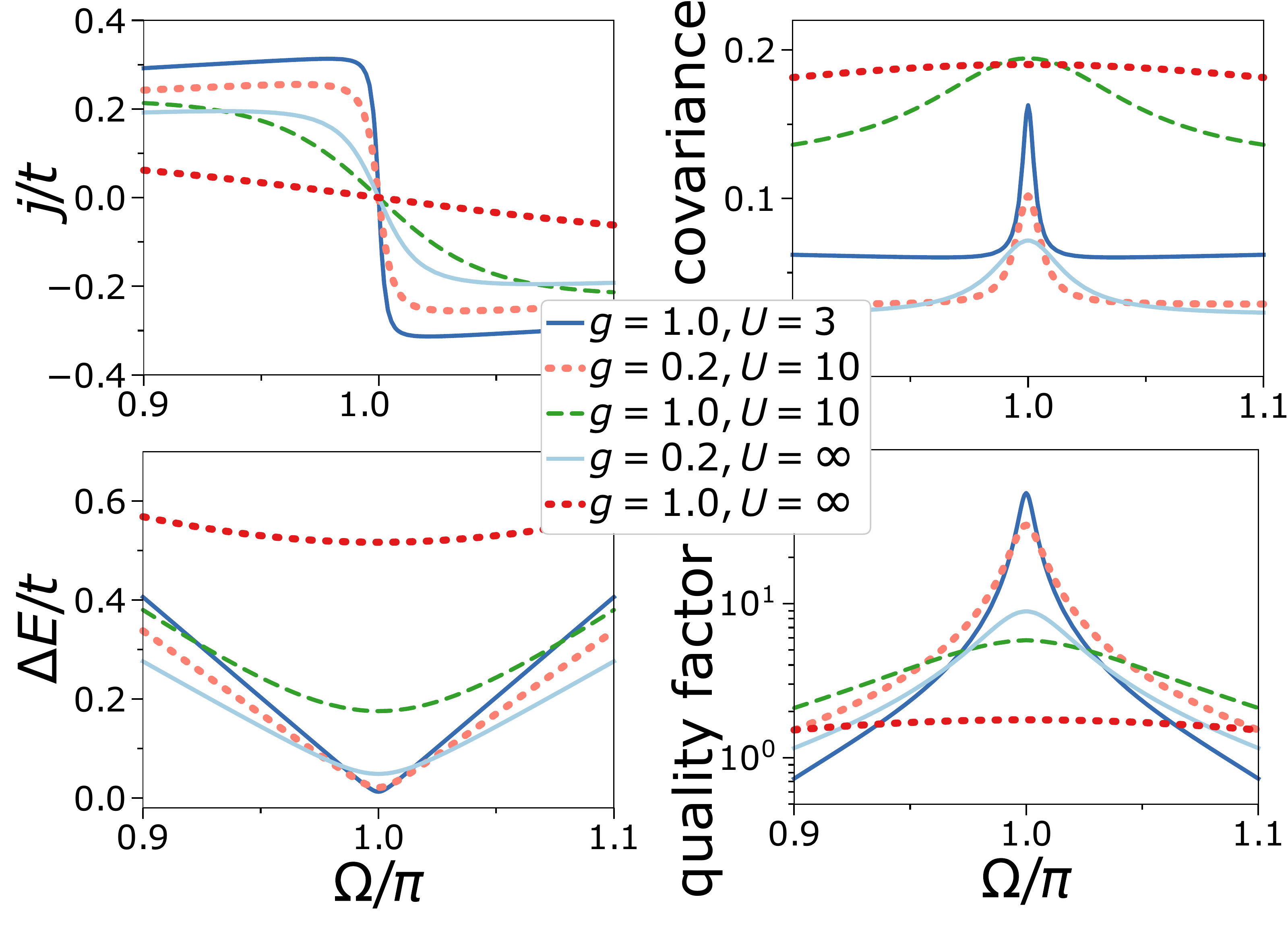}}
	\caption{Two coupled rings with 9 sites each and 9 particles threaded by the same total flux ${\Omega=\phiA L=\phiB L}$ and ${w=0}$ for different values of interaction $U/t$ and inter-ring coupling $g/t$. {\bf a)} Average current in lab-frame in either ring. {\bf b)} Covariance (a measure of correlation) of currents in ring A and B ${\text{cov}=\mean{j_\text{A}j_\text{B}}-\mean{j_\text{A}}\mean{j_\text{B}}}$. {\bf c)} energy gap $\Delta E$ between ground state and first excited state. {\bf d)} Quality factor $Q$, which is the ratio of energy difference of second excited state and first excited state, and $\Delta E$. ${Q>1}$ defines a good qubit. Further discussion of the qubit parameters is found in \protect\cite{aghamalyan2015coherent}.}
	\label{CoRotPhaseslip}
\end{figure}
Here, we report that 
${\phiA L+\phiB L=2\pi}$, ${\phiA L \approx \pi}$ results defining a working point for a two level system as well. 
In Fig.\ref{CoRotPhaseslip} the average current in either ring, current covariance, energy gap and quality factor (as defined in \cite{aghamalyan2015coherent}) is studied in dependence of  ${\Omega=\phiA L=\phiB L}$, for various values of interaction $U/t$ and inter-ring coupling $g/t$. In Fig.\ref{CoRotPhaseslip}a, we observe a change from small to large average current across the optimal working point ${\Omega=\pi}$. At that point in Fig.\ref{CoRotPhaseslip}b, the current-current correlation (the so-called covariance) displays a positive maximum, which indicates that the direction of the currents in the both rings are positively correlated Thus, the two levels are now provided by co-propagating currents in the two rings. 
This is in contrast the other working point, where the flux and current in each ring has opposite sign.

%

\section{Artificial gauge field}
The artificial gauge field can be generate by following different protocols: two photon Raman transitions, suitable gradient laser intensity, steering and suitable shaking of the potentials. 

For the potential shaking: If the shaking is fast compared to the Hamilton dynamics, it will generate an effective, time-independent tunneling parameter. For sinusoidal shaking, it is possible to control the sign of the tunneling. When the driving breaks time-reversal symmetry, complex tunneling constants are possible, which can be engineered into artificial gauge fields\cite{struck2012tunable}. 
Artificial gauge fields also have been realized using laser assisted tunneling\cite{atala2014observation,dalibard2011colloquium}. All such protocols can be be applied to the two coupled rings if the two rings are sufficiently distant apart. 

Another option to construct an artificial gauge field is the rotation of the lattice. Experimentally this can be realized by stirring the condensate, e.g. by a moving a barrier through the ring with a constant velocity. 
\revA{This method has been experimentally realized with a single continuous ring\cite{wright2013driving}. For two rings, the barrier in each ring would need opposite velocity in each ring. Experimentally, the difficulty lies in applying the barriers with opposite velocity in each ring separately, without disturbing the other. This could be done by separating the ring potentials at large enough distance, such that a laser for each individual barrier can be independently realized in each ring without influencing the other ring. After preparing the desired winding state, the rings are brought slowly into contact. For very high flux, this method may not work well.

We now proceed to describe the creation of the artificial gauge field in a single ring.} In the rotating frame, the Hamiltonian of a particle in a ring becomes (while neglecting the barrier contribution)
\begin{equation}
\mathcal{H}=\frac{\op{p}^2}{2m}-\Omega_\text{rot}\op{L_z}~,
\end{equation}
with $\op{p}$ the momentum operator, $\op{L_z}$ the angular momentum operator and $\Omega_\text{rot}$ the angular velocity of the stirring by the external force. For a one-dimensional ring, this equation can be rewritten with an effective gauge field $qA(R)$ as
\begin{align}
\mathcal{H}={}&\frac{(\op{p}-qA)^2}{2m}-V_\text{centr}\\
qA={}&\Omega_\text{rot} Rm\\
V_\text{centr}={}&\frac{\Omega_\text{rot}^2R^2m}{2}
\end{align}
where $V_\text{centr}(R)$ is centrifugal potential. The resulting force of the gauge field is the Coriolis force. 

Akin to the magnetic flux threading a superconducting loop, we define the Coriolis flux threading the system
\begin{equation}
\Omega=\frac{q}{\hbar}\oint_C\vc{A}\diff{\vc{r}}=2\pi\frac{ m\Omega_\text{rot} R^2}{\hbar}~,
\end{equation}
where $\omega$ is the angular velocity of the barrier, $m$ the mass of the particles and $R$ the radius of the ring. 
In a ring, the angular momentum of a particle is quantized in integer number $n$ due to the periodic boundary condition and given by 
\begin{equation}
k_n=\frac{2\pi n }{L_s}~,
\end{equation}
with $L_s=2\pi R$ the length of the ring.
The energy of the particle is then
\begin{equation}
E=\frac{(\hbar k_n-qA)^2}{2m}=\frac{\hbar^2}{2m}\left(\frac{2\pi}{L_s}(n-\frac{\Omega}{2\pi})\right)^2
\end{equation}
The ground state energy of a particle in a ring changes with flux $\Omega$ with a period of $2\pi$. The angular velocity of the condensate in the non-rotating frame is given by $\omega=\frac{\hbar k}{mR}$ and is quantized in integer multiples of 
\begin{equation}
\omega_0=\frac{\hbar}{mR^2}~.
\end{equation}
For a lattice system, the ring is discretized into $L$ sites and the intra-ring tunneling becomes
\begin{equation}
\mathcal{H}=\sum_{m=1}^{L}\left(-t\expU{i\Omega/L}\cn{a}{m}\an{a}{m+1}+\text{h.c.}\right)~.
\end{equation}
The phase shift for tunneling between neighboring sites is given by
\begin{equation}
\phi=\frac{2\pi m\omega  R^2}{\hbar L}=\frac{2\pi mvR}{\hbar L}~,
\end{equation}
where $v$ the velocity of the rotation. 
For the two ring setup proposed in \cite{amico2014superfluid} with a ring radius $5~\mu\text{m}$ and loaded with Rubidium atoms, the condensate rotates in multiples of the frequency $f_0=4.73\text{Hz}$.

\section{Time of flight}\label{AppendixTOF}
The atoms are released from the trap and expand freely for a certain amount of time. The momentum distribution $n(\vc{k})$ corresponds then to the particle density of the expanded gas. 
We calculate it with the Fourier transform of the one-body density matrix $\rho_{(1)}(\vc{x},\vc{x'})=\langle\hat\psi^\dagger(\vc{x})\hat\psi(\vc{x'})\rangle$ and get
\begin{equation}
\label{TOF1}
n(\vc{k})=\int\diff{x}\int\diff{x'}\langle\hat\psi^\dagger(\vc{x})\hat\psi(\vc{x'})\rangle\expU{i\vc{k}(\vc{x}-\vc{x'})}~.
\end{equation}
To account for the lattice structure, we express the equation in terms of Wannier functions $w_j(\vc{x})=w(\vc{x}-\vc{x}_j)$, which are localized at the location of lattice site $\vc{x}_j$. The boson operator becomes $\psi(\vc{x})=\sum_{j}w_j(\vc{x})\an{c}{j}$, where $\an{c}{j}$ is the annihilation operator of the Bose-Hubbard operator for site $j$ and the sum runs over all sites of both rings. With this, Eq. \ref{TOF1} is recast to
\begin{equation}
n(\vc{k})=\abs{\tilde w(\vc{k})}^2\sum_{l,j}\expU{i\vc{k}(\vc{x}_l-\vc{x}_j)}\langle\cn{c}{l}\an{c}{j}\rangle~,
\end{equation}
where $\tilde w(\vc{k})$ is the Fourier transform of the Wannier function.

\end{document}